\documentclass[aps]{revtex4}
\usepackage{amsmath,amssymb,graphicx,float}
\begin{document}
\title{Currents and correlations in Luttinger
liquids and carbon nanotubes at finite temperature and size:  a
bosonization study}
\author{J.-S. Caux$^{1,2,3}$, A. L{\'o}pez$^{1,}$\footnote{Permanent Address:
 Centro At{\'o}mico Bariloche - Instituto Balseiro, 8400 Bariloche, Argentina. } and D. Suppa$^{1}$}
\affiliation{$^1$Theoretical Physics, University of Oxford, 1 Keble Road,
Oxford, OX1 3NP, UK \\
$^2$ All Souls College, Oxford, OX1 4AL, UK \\
$^3$ ITFA, University of Amsterdam, Valckenierstraat 65, 1018 XE
Amsterdam, The Netherlands} 
\date{\today}
\begin{abstract}
We consider problems of one dimensional interacting fermions confined to a finite
size, multichannel geometry.  Concentrating on Luttinger liquids and
carbon nanotubes, we use nontrivial boundary conditions to represent
the effect of external leads, and apply our framework to transport
problems in a Josephson junction setup.  We present an exact
computation of all correlation functions, including finite-size and
temperature effects, for two sets of solvable boundary conditions.  In
all cases, we compute physical quantities like the Josephson current
and the pairing order parameter profile.
\end{abstract}
\maketitle

%%%%%%%%%%%%%%%%%%%%%%%%%%%%%%%%%%%%%%%%%%%%
%%%%%%%%%%%%%%%%%%%%%%%%%%%%%%%%%%%%%%%%%%%%
\section{Introduction}
\label{introduction}

It is now common knowledge that one-dimensional interacting electronic
systems possess metallic phases that are fundamentally different from
those of their higher-dimensional brethren, which are for most
practical purposes extremely well described by adaptations and
refinements of Fermi liquid theory.  The Luttinger liquid has become
the paradigm for the former systems:  the quasiparticle pole is no
more, and instead one finds only collective spin and charge
excitations living around two Fermi points.  The most striking
consequence of interactions are first that spin and charge excitations are not
bound to one another anymore, and can thus travel at different
velocities.  Moreover, charge fractionalization can occur, whereby
``fundamental'' quantities like the electron's charge can be split in
many pieces (for a review, see {\it e.g.} \cite{AlexeiBook}).

Experimental realizations of Luttinger liquids range from edge states
in the fractional quantum Hall effect \cite{ChangPRL77}
(in this case, a chiral Luttinger liquid is found), quantum wires in
semiconductor heterostructures \cite{YacobyPRL77}, and perhaps most
notably in single-walled carbon nanotubes (SWNT)
\cite{BockrathNature397,EggerPRL79,KanePRL79}.  The latter can be 
used to perform electrical transport experiments using different
geometries, for example in
junctions manipulated by  
mechanical means \cite{TansNature386,YaoNature402,PostmaPRB62}.  
``Kinks'' in the nanotubes or crossings of
nanotubes can be produced by scanning tunneling microscopes (STM)
tips, and these in turn provide realizations of backscattering
impurities embedded in a Luttinger liquid or of connections through
tunnel junctions. These junctions can act as rectifying diodes at room temperature,
thus opening up a route towards the construction of
nanoscale devices of all sorts.  An example of recently proposed
applications is that of an ``entangler'', whereby two nanotubes
coupled to a superconductor are expected to produce physically
separated entangled pairs of excitations \cite{RecherPRB65,BenaPRL89, Bouchiat02}.

Single impurities embedded in Luttinger liquids have attracted a lot
of theoretical interest since the early work of Kane and Fisher
\cite{KanePRB46}.  In particular, many approaches have been devised
based on boundary conformal field theory (CFT) \cite{WongNPB417} and
boundary integrability \cite{FendleyPRB52,SaleurLectures}.  The trick
here is to ``fold'' 
the system around the impurity, thereby replacing it with an effective
boundary whose effects scale either to the weak- or strong-coupling
regime depending on the nature of the bulk interactions in the
system.  For example, for a backscattering impurity in a Luttinger
liquid, repulsive interactions enhance the backscattering amplitude,
and therefore suppress the conductance.

Another extremely interesting situation to consider is that of a
Luttinger liquid on a finite interval.  This class of problems is a
form of generalization of the previous single impurity ones, in the
sense that we now have two different boundaries to deal with.  One
thus imagines electrons being trapped by either a direct physical cut
of their support (maybe an ended nanotube), or prevented from
propagating further by the presence of an excitation gap, 
maybe a superconducting gap, or voltage gates. 
We consider here the problem
of a Josephson junction realized by bridging two superconductors with
either quantum wires or carbon nanotubes.  Experimental attempts at
realizing this device can be found in \cite{KasumovScience284}.
Theoretically, this problem has been addressed in \cite{MaslovPRB53},
where the Josephson current for ``perfect'' contacts was computed
(here, by perfect, we mean a contact such that only Andreev reflection
occurs;  a poor contact would be one such that most of the reflection
processes that occur would be normal reflection ones).  A
perturbative analysis of the problem was presented in
\cite{FazioPRB53}, and an effective model amenable to boundary CFT was
used in \cite{AffleckPRB62}.  For certain cases, the problem is an
integrable quantum field theory, and this was treated in
\cite{CauxPRL88}.  However, only one particular (nontrivial) value of the
interaction parameter can be fully solved, the others leading to
rather tedious complications with the analytical properties of the
thermodynamic Bethe Ansatz, which are yet to be fully resolved.  One
common factor of all these studies, however, is that they were devoted
to single-channel Luttinger liquids. 

In the case of nanotubes, the number of channels available for
conduction is however by construction greater than one.  
Moreover, if one wants to
describe interesting devices like the entangler mentioned above, or 
maybe even things like multiwall carbon nanotubes \cite{MWCN}, it is
important to have a theory for multichannel situations.  
It is very tempting to think that the one-dimensionality of the system
opens the way to a full nonperturbative solution of transport problems
of this class, as it can do for the single-channel case.  However,
from basic factorized scattering arguments, it is possible to show
\cite{Delfino} that multichannel theories nontrivially coupled only at
boundaries but not in the bulk, do not seem to have enough
flexibility to be integrable (that is, only a trivial scattering
matrix can satisfy the bootstrap equations).  This holds, of course,
if one considers having no boundary degrees of freedom.  Relaxing this
might lead to some other integrable theories, but not ones which can
be directly applied to the problem at hand.  One is
therefore ultimately led straight back into the domain of perturbative
computations as the only fully controllable approach, and most of what
we will present here shall be based on a minimal use of it (that is,
we solve all that we can exactly, using perturbation theory only when
strictly needed).  

Our plan is therefore to treat the general problem of a conducting
multichannel system in a finite size domain, between two boundaries having
nontrivial effects on the physics (but no independent degrees of
freedom of their own).  The noninteracting and interacting cases are
treated in fundamentally different manners.
The noninteracting problem is straightforward
to solve thoroughly directly in the fermionic language, providing in
some sense an adaptation of the 
well-known work of Blonder, Tinkham and Klapwijk 
\cite{BlonderPRB25} to two-boundary problems.  The partition function
can be computed exactly, which means that thermodynamic properties and
correlation functions can be obtained exactly (for the case of
correlation functions, we only solve the interacting case, since the
noninteracting ones can be recovered easily by putting all interaction
parameters to zero).  

Including interactions, however, takes us to a 
different set of problems.  First of all, the presence of appropriate
Coulomb interactions renders the fermionic description untractable.
Fortunately, however, we can use bosonization to perform our
computations.  This, as is well-known, allows one to treat interactions
nonperturbatively.  In the bulk, correlators are rather easily
computed.  Remember, however, that we are here dealing with
two-boundary problems.  This involves treating Luttinger liquids and
carbon nanotubes in a 
finite size at finite temperatures, for which we present a computation
of all multipoint correlation functions around simple fixed points
represented by tractable boundary conditions.  The scaling of the
correlators then depends on whether the operators sit away from or
near to the boundaries:  the scaling near the boundaries differs from
that in the bulk, and depends on the specifics of the boundary
conditions that are in use.  We venture to treat this problem in all
generality at the level of bosonization.  For our specific problems, 
we set up the perturbative formalism, and
present computations for physical transport quantities like the
Josephson current in different situations, as well as discussions
about the pairing order parameter profile.  

The plan of the paper is as follows.  In section 2, we treat the
noninteracting problem in fermionic language, and compute
finite-temperature transport properties using canonical mode expansion
methods.  In section 3, we consider the interacting case.  We consider
separately the cases of boundary-coupled Luttinger liquids, and
nanotubes.  For each of these, we discuss separately the cases of
normal and Andreev boundary conditions, representing ``bad'' and
``good'' coupling to the superconductors (again, in the sense
described above in terms of normal versus Andreev reflection
amplitudes;  the microscopics of this is a rather tricky experimental
issue).  The physical results are 
listed in this part of the paper, and summarized in the conclusion. 
Important formulas for the bosonization, including
computations of correlators at finite temperature in a finite-size
geometry are collected in Appendix A.  In Appendix B, we provide
some definitions of (multivariable) $\theta$-functions appearing in
our formulas for the correlation functions.  Appendix C is devoted to
some notes on the real-time, finite-temperature perturbative formalism
we used.  

%%%%%%%%%%%%%%%%%%%%%%%%%%%%%%%%%%%%%%%%%%%%
%%%%%%%%%%%%%%%%%%%%%%%%%%%%%%%%%%%%%%%%%%%%

\section{Noninteracting case}

In this section, we concentrate on the case where interactions are
absent in the bulk of the system.  The main advantage of this is that 
the calculation of thermodynamic quantities and of
correlators can be performed exactly using relatively straightforward
methods.  We choose to always work in a real-time formalism, even in
the presence of finite temperatures.  This allows an immediate
adaptation of our framework to time-dependent perturbations (although
we do not consider them here, they are in principle easy to address
using our formulas, bearing the notes in Appendix C in mind).  
Moreover, the definitions and results obtained in this section, 
in particular the treatment of boundary phenomena,  will
prove to be a good starting point for other problems later on,  when
we will consider other physical cases with 
interactions in the bulk.  Our intention is to provide a flexible
framework for building theories of {\it e.g.} transport through
one-dimensional channels, in many different types of setups.  Recent
advances in experimental realizations of nanostructures provide good
motivation for such types of theories.

Our starting point is a theory for $N$ decoupled spinful channels of
fermions living within a one-dimensional channel of finite size, $x
\in [0,R]$, which we often call the fundamental domain.  
The spin-$1/2$ Fermi fields are labeled by channel $i = 1, ..., N$ and
spin indices, 
$\Psi_{\sigma}^j (x)$.  
The canonical equal-time anticommutation relations for the fermions
are (for $x, x' \in ]0,R[$)
\begin{eqnarray}
\left\{ \Psi_{\sigma}^{j} (x,t), \Psi_{\sigma'}^{j' \dagger} (x',t)
\right\} = \delta_{\sigma \sigma'} \delta^{j j'} \delta (x - x').
\end{eqnarray}
In order to write proper mode expansions for our fermions, we have to
be careful with what happens in the presence of the two boundaries.  The
crucial thing to preserve is the independence of $\Psi_{\sigma}^j (x)$
at both boundaries:  we do not want to identify fields at $x = 0$ and
fields at $x = R$, as one would normally do using a finite-size setup
on a ring.  We here have to start by unfolding the fields, and then
requiring some form of periodicity.  The way we do this is by 
extending the definitions of the fields to the
whole real axis by reflecting the original field at both ends of the
system, {\it i.e.} by taking
\begin{eqnarray}
\Psi_{\sigma}^j (-x) = \Psi_{\sigma}^j (x), \hspace{1cm}
\Psi_{\sigma}^j (R-x) = \Psi_{\sigma}^j (R+x).
\label{extensions}
\end{eqnarray}
This immediately shows that the proper periodicity of the fields is
$2R$, and not $R$ as we would have obtained in a closed ring of
circumference $R$.  

Each channel has a value of the Fermi wavevector $k_F^j$ and
Fermi velocity $v^j$ which we take as external parameters.
This gives us the flexibility for example to treat the cases where the
different channels are put at different external voltages \cite{Egger-Grabert}.  
In the simplest case, the
relationship between the 
external voltage $V^j$ applied to channel $j$ and the Fermi parameters
is $k_F^j = k_F^0 + \frac{V^j}{2v^j}$, where $k_F^0$
is the Fermi wavevector at 
half-filling.  This step in not crucial at this stage, as
most of our formulas will comprise boundary backscattering potentials
representing the voltages of leads, so we can simply keep the voltages
explicitly instead of reabsorbing them.  For transport, only voltage
differences matter.  
The Fermi fields are then linearized as usual in terms of chiral left- and
right-movers as
\begin{eqnarray}
\Psi_{\sigma}^j (x) = e^{i k_F^j x} \Psi_{R \sigma}^j (x) +
e^{-i k_F^j x} \Psi_{L \sigma}^j (x).
\end{eqnarray}
with canonical equal-time anticommutation relations
\begin{eqnarray}
\left\{ \Psi_{L \sigma}^{j} (x,t), \Psi_{L \sigma'}^{j' \dagger} (x',t)
\right\} = \frac{1}{2} \delta_{\sigma \sigma'} \delta^{j j'} \delta (x
- x') 
\label{fermioncancomm}
\end{eqnarray}
and a similar equation for the right movers (the left and right fields
anticommute in the fundamental domain;  note the factor of a
half, which is usual for chiral fields \cite{FloreaniniPRL17}).  
In terms of left- and right-movers, the extensions (\ref{extensions})
imply the analytical continuations of chiral fields
\begin{eqnarray}
\Psi_{L \sigma}^j (-x) = \Psi_{R \sigma}^j (x), \hspace{1cm} 
\Psi_{L \sigma}^j (R-x) = e^{2i k_F^j R} \Psi_{R \sigma}^j (R+x),
\end{eqnarray}
which represent a full mapping between the left- and right-movers in
the theory (that is, we can use one or the other, as long as we choose
the right periodicity for getting the proper mode expansions).  The
appropriate periodicity of the fields then becomes
\begin{eqnarray}
\Psi_{L \sigma}^j (2R + x) = e^{2i k_F^j R} \Psi_{L \sigma} (x),
\hspace{1cm} \Psi_{R \sigma}^j (2R + x) = e^{-2i k_F^j R} \Psi_{R
\sigma} (x).
\end{eqnarray}
The above formulas in turn imply what we term the normal boundary
conditions at the left and right ends of the system, representing the
fact that excitations hitting the boundary are simply normal-reflected
back into the bulk of the system:
\begin{eqnarray}
\Psi_{L \sigma}^j (0) = \Psi_{R \sigma}^j (0), \hspace{1cm}
\Psi_{L \sigma}^j (R) = e^{2i k_F^j R} \Psi_{R \sigma}^j (R).
\label{normalbcs}
\end{eqnarray}
The important point to realize is that these boundary conditions
represent a link between the fields of the theory:  using
them, all the fields of the theory can be described either by
left- and right-movers in the fundamental domain, or using only
e.g. left-movers, defined over an interval twice as big as the
fundamental domain.  This set of boundary conditions is used
throughout (half) of the paper, and represents a basis from which
bosonization can be done exactly.  They should be kept closely in mind
by the reader.  

The action describing the left- and right-movers with normal boundary
conditions is composed of a noninteracting bulk part and a specific
boundary part which is chosen such as to canonically impose the normal
boundary conditions (\ref{normalbcs}):   
\begin{eqnarray}
S_{norm} = S^{(bulk)}_0 + S^{(bdry)}_{norm}.
\label{action}
\end{eqnarray}
The noninteracting bulk action is most conveniently written in terms
of a symmetric form of chiral left- and right-movers:
\begin{eqnarray}
S^{(bulk)}_0 = \int dt \int_0^R dx \sum_{j=1}^N \sum_{\sigma = \uparrow,
\downarrow} \left
[ \frac{1}{2} 
\Psi_{L \sigma}^{j \dagger} i (\partial_t - v^j \partial_x) \Psi_{L
\sigma}^j 
+ \frac{1}{2} {\Psi_{L
\sigma}^j} i (\partial_t - v^j \partial_x) \Psi_{L
\sigma}^{j \dagger} + \right. \nonumber \\
\left. + \frac{1}{2} \Psi_{R
\sigma}^{j \dagger} i (\partial_t + v^j \partial_x) \Psi_{R \sigma}^j
+ \frac{1}{2} \Psi_{R
\sigma}^j i (\partial_t + v^j \partial_x) \Psi_{R
\sigma}^{j \dagger} \right].
\end{eqnarray}
The boundary action imposing normal boundary conditions is
\begin{eqnarray}
S^{(bdry)}_{norm} = \int dt \sum_j \sum_{\sigma} i \frac{v^j}{2}
\left[\Psi_{L\sigma}^{j \dagger} (0) \Psi_{R \sigma}^j (0) -
\Psi_{R \sigma}^{j \dagger} (0) \Psi_{L \sigma}^j (0) 
- \right. \nonumber \\
\left. - e^{2ik_F^j R} \Psi_{L
\sigma}^{j \dagger} (R) \Psi_{R \sigma}^j (R) 
+ e^{-2ik_F^j R} \Psi_{R
\sigma}^{j \dagger} (R) \Psi_{L \sigma}^j (R) \right].
\label{fermionicnormalBCs}
\end{eqnarray}
This theory is very simple:  its partition function and correlators
are straightforward to obtain, even at finite temperature.  One can
easily verify that the equations of motion at the boundaries coincide
with the boundary conditions (\ref{normalbcs}).

What really interests us, however, is to couple the system to external
leads.  One of the most interesting configurations that we can
consider is the one where the system is in a Josephson junction
geometry, whereby the left and right ends are in contact with two
superconductors having different order parameters.  On general
grounds, one then expects currents to flow through the system even in
the absence of voltage bias, part of the 
phenomena called the Josephson effects \cite{JosephsonPL1}.  

We therefore consider coupling our system to two external
superconductors with bulk gaps well above the excitation energies
existing within the conducting channel.  In particular, this means
that quasiparticle penetration within the superconductors from the
channel are basically non-existent, and that the excitations are
either normal- or Andreev-reflected immediately back into the channel by the
contacts with the superconductors.  An effective theory is therefore
obtained by integrating out the superconductors themselves
\cite{AffleckPRB62}, leaving behind effective boundary actions taking
the form of BCS-like couplings at the ends of the system.  Therefore,  
we perturb the free action by the additional boundary actions
\begin{eqnarray}
S^{(bdry)}_{L} = - \frac{1}{2} \int dt \sum_{i,j = 1}^N \left
[ \Delta^{ij}_L 
\Psi_{\uparrow}^{i \dagger} (0) \Psi_{\downarrow}^{j \dagger} (0) +
\Delta^{ij *}_L \Psi_{\downarrow}^j (0) \Psi_{\uparrow}^i (0) +
V^{ij}_L \sum_{\sigma} \Psi_{\sigma}^{i \dagger} (0) \Psi_{\sigma}^j
(0) \right], \nonumber \\ 
S^{(bdry)}_{R} = - \frac{1}{2} \int dt \sum_{i,j = 1}^N \left
[ \Delta^{ij}_R 
\Psi_{\uparrow}^{i \dagger} (R) \Psi_{\downarrow}^{j \dagger} (R) +
\Delta^{ij *}_R \Psi_{\downarrow}^j (R) \Psi_{\uparrow}^i (R) +
V^{ij}_R \sum_{\sigma} \Psi_{\sigma}^{i \dagger} (R) \Psi_{\sigma}^j
(R) \right]
\end{eqnarray}
The boundary BCS couplings $\Delta_{L,R}$ are effective couplings
coming from the integrating out procedure detailed above (we refer the
reader to the discussion in paper \cite{AffleckPRB62}, where this is
presented in all necessary details).  It is very
important not to confuse them with the bulk couplings of the original
superconductors:  the boundary pairings contain information about the
quality of the contact, the density of states of the superconductor at
the junction, etc.  We take them as external parameters.  Their
phase, however, can be reasonably supposed to be that of the original
superconductors at low energies.  

The one advantage of this form of the boundary perturbation is that it
preserves the bilinear form of the action as far as the fermions are
concerned.  The theory can therefore again be solved exactly.
Varying the action yields the interpolating boundary conditions
\begin{eqnarray}
\Psi_{L \sigma}^i (0) -\frac{i}{v^i} \sum_j \sum_{\sigma'} \left
[ \Delta^{ij}_L \epsilon_{\sigma \sigma'} \Psi_{L \sigma'}^{j \dagger}
(0) + V^{ij}_L \delta_{\sigma \sigma'} \Psi_{L \sigma'}^j (0) \right]
= \nonumber \\
= \Psi_{R \sigma}^i (0) + \frac{i}{v^i} \sum_j \sum_{\sigma'} \left
[ \Delta^{ij}_L \epsilon_{\sigma \sigma'} \Psi_{R \sigma'}^{j \dagger}
(0) + V^{ij}_L \delta_{\sigma \sigma'} \Psi_{R \sigma'}^j (0) \right],
\nonumber \\
e^{-i k_F^i R} \Psi_{L \sigma}^i (R) 
+\frac{i}{v^i}  \sum_j \sum_{\sigma'} \left
[ \Delta^{ij}_R \epsilon_{\sigma \sigma'} e^{i k_F^j R} 
\Psi_{L \sigma'}^{j \dagger}
(R) + V^{ij}_R \delta_{\sigma \sigma'} e^{-i k_F^j R}
\Psi_{L \sigma'}^j (R) \right]
= \nonumber \\
= e^{ik_F^i R} \Psi_{R \sigma}^i (R) 
- \frac{i}{v^i} \sum_j \sum_{\sigma'} \left
[ \Delta^{ij}_R \epsilon_{\sigma \sigma'} e^{-i k_F^j R} 
\Psi_{R \sigma'}^{j \dagger}
(R) + V^{ij}_R \delta_{\sigma \sigma'} e^{i k_F^j R}
\Psi_{R \sigma'}^j (R) \right].
\label{boundaryconditions}
\end{eqnarray}
where $\epsilon_{\uparrow \downarrow} = - \epsilon_{\downarrow
\uparrow} = 1$.  These boundary conditions allow us to move
continuously between various extreme cases, where things simplify
considerably.  Besides the normal boundary conditions above, we also
can tune our boundary parameters to achieve what we will call Andreev
boundary conditions:  taking
\begin{eqnarray}
V^{ij} = 0, \hspace{1cm} \Delta^{ij}_L = 
v^i \delta_{ij}, \hspace{1cm} \Delta^{ij}_R = v^i e^{i
\chi^i} \delta_{ij},
\end{eqnarray} 
the Andreev boundary conditions read explicitly
\begin{eqnarray}
\Psi_{L \sigma}^j (0) = i  \epsilon_{\sigma \sigma'}
\Psi_{R \sigma'}^{j \dagger} (0), \hspace{1cm}
\Psi_{L \sigma}^j (R) = -i e^{i \chi^j} \epsilon_{\sigma \sigma'}
\Psi_{R \sigma'}^{j \dagger} (R).
\label{Andreevbcs}
\end{eqnarray}
The two cases of normal and Andreev boundary conditions are those for
which one can compute thermodynamic quantities exactly in the
interacting case, and we urge the reader to keep them in mind in what
follows.  Away from these, one has to rely on some form of 
perturbation theory.

In the absence of interactions, however, the exact solution of the
theory is obtained by simple considerations.  Since the fermions are
free chiral fields in the bulk, we can write the mode expansions
\begin{eqnarray}
\Psi_{L \sigma}^j (x,t) = \int \frac{dk}{2\pi} e^{-i k (x + v^j t)}
\Psi_{L \sigma}^j (k), \hspace{1cm} 
\Psi_{R \sigma}^j (x,t) = \int \frac{dk}{2\pi} e^{i k (x - v^j t)}
\Psi_{R \sigma}^j (k).
\label{modeexpansions}
\end{eqnarray}
Substituting (\ref{modeexpansions}) in (\ref{boundaryconditions}) and
defining the Nambu spinors 
\begin{eqnarray}
\Psi_{L,R} (\omega) = \left( 
\begin{array}{c}
\Psi_{L,R \uparrow}^1 (\frac{\omega}{v^1}) \\ ... \\ \Psi_{L,R
\downarrow}^{1 \dagger} (-\frac{\omega}{v^1}) \\ ... 
\end{array} \right)
\end{eqnarray}
we get the continuity equations
\begin{eqnarray}
\left[ 1 - i {\bf v}^{-1} {\bf M}_L \right] \Psi_L (\omega) = 
\left[ 1 + i {\bf v}^{-1} {\bf M}_L \right] \Psi_R (\omega), \nonumber \\
\left[ 1 + i {\bf v}^{-1} {\bf M}_R \right] 
 e^{-i {\bf \sigma}_z \otimes {\bf k}_F
R - i {\bf 1} \otimes {\bf v}^{-1} \omega R} 
\Psi_L (\omega) = 
\left[ 1 - i {\bf v}^{-1} {\bf M}_R \right] e^{i {\bf \sigma}_z
\otimes {\bf k}_F 
R + i {\bf 1} \otimes {\bf v}^{-1} \omega R} 
\Psi_R (\omega),
\end{eqnarray}
where ${\bf v} = \textbf{diag} (v^i)$, and
the boundary matrices are given by
\begin{eqnarray}
{\bf M}_{L,R} = 
\left(\begin{array}{cc}
{\bf V}_{L,R} & {\bf \Delta}_{L,R} \\
{\bf \Delta}^*_{L,R} & - {\bf V}^*_{L,R}
\end{array} \right).
\end{eqnarray}
(note:  we use here a Bogoliubov-de Gennes notation, with each entry
of the 2x2 matrices above being an NxN matrix, where N is the number
of channels.  We order the tensor products as (BdG space) $\otimes$
(channels space)).
The quantization condition is then obtained by the requirement that a
given eigenstate energy $\omega$ fulfills both continuity equations.
This implies 
\begin{eqnarray}
\det \left[ 1 + F(\omega) \right] = 0, 
\end{eqnarray}
with matrix 
\begin{eqnarray}
F (\omega) = - \left[ 1 - i {\bf v}^{-1} {\bf M}_L \right]^{-1} 
\left[ 1 + i {\bf v}^{-1} {\bf M}_L \right] e^{-i {\bf \sigma}_z
\otimes {\bf k}_F 
R - i {\bf 1} \otimes {\bf v}^{-1} \omega R} \times \nonumber \\
\times 
\left[ 1 - i {\bf v}^{-1} {\bf M}_R \right]^{-1} 
\left[ 1 + i {\bf v}^{-1} {\bf M}_R \right] e^{-i {\bf \sigma}_z
\otimes {\bf k}_F 
R - i {\bf 1} \otimes {\bf v}^{-1} \omega R}.
\end{eqnarray}
Noting that we can write
\begin{eqnarray}
\left[ 1 - i {\bf v}^{-1} {\bf M} \right]^{-1} 
\left[ 1 + i {\bf v}^{-1} {\bf M} \right] = e^{2i \arctan ({\bf
v}^{-1} {\bf M})},
\label{bdrymatrix}
\end{eqnarray}
we see that $F^{\dagger} = F^{-1}$, so 
all solutions to the quantization equation are such that $\omega \in
\mathbb{R}$. 
 
The ground-state energy of the system is then given simply by the sum
over a Fermi sea, which we can write as a contour integral:
\begin{eqnarray}
E_0 =  \sum_{\omega < 0} \omega = \frac{1}{2\pi i} \int_C d \omega
\omega 
\frac{\frac{\partial \det[1 + F(\omega)] }{\partial \omega}}{\det[1 +
F(\omega)]} = -
\frac{1}{2\pi i} \int_C d \omega \ln \det [1 + F(\omega)]
\end{eqnarray}
where the contour $C$ circles all the poles of $\det [1+F(\omega)]^{-1}$ on
the negative real axis, in a clockwise direction.
Using complex plane manipulations, we can write
\begin{eqnarray}
E_0 = -\frac{1}{2\pi i} \int_{-i \infty}^{i \infty} d \omega
\ln \det [ 1 + F(\omega)] = -\frac{1}{2\pi} \int_0^{\infty} d \eta
\ln \det \left([ 1 + F(-i \eta)][ 1 + F(i \eta)] \right) = \nonumber
\\
= -\frac{1}{2\pi} \int_0^{\infty} d \eta
\ln \det \left([ 1 + F(-i \eta)][ 1 + F^{-1}(i \eta)] \right) + G_L + G_R
\end{eqnarray}
where the first term is now manifestly convergent, and the additional
contributions are the boundary intensive energies, which we ignore
from now on since they have no influence on the physics that we are after.

We can now introduce finite temperatures by inserting
temperature-dependent occupation numbers for all states, and
recalculating the sum.  We find, for the equilibrium distribution,
\begin{eqnarray}
\ln Z = \int_{-\infty}^{\infty} d \omega D(\omega) \ln \left[ 1 + e^{-\beta
|\omega|} \right] - \beta E_0
\end{eqnarray}
where the thermal density of states is given by the expression
\begin{eqnarray}
D(\omega) = \frac{1}{\pi} \lim_{\delta \rightarrow 0} \Im
\frac{\partial}{\partial \omega} \ln \det \left[ 1 + F (\omega - i
\delta) \right]
\end{eqnarray}
where $\Im$ denotes the imaginary part.
A more useful expression for numerical purposes is obtained by closing
the contour around the poles of the Fermi distribution.  We then get
\begin{eqnarray}
\ln Z = \sum_{n=0}^{\infty} \mbox{Tr} \ln \left[[ 1 + F(-2\pi (n+1/2)
i/\beta)] [1 + F^{-1} (2\pi (n+1/2) i/\beta)] \right].
\label{finiteTZ}
\end{eqnarray}
Although the sum cannot be taken in closed form for general boundary
couplings, higher-order terms become exponentially suppressed at a
given finite temperature.  For zero temperature, the sum becomes an
integral reproducing the formula above.

The crucial point to bear in mind here is that we now have an
expression for the partition function, for finite size and finite
temperatures, for an arbitrary set of boundary parameters.  These,
hidden in the matrices above (which contain the boundary matrices), 
allow one in principle to compute many thermodynamic quantities in the
system by using simple parameter derivatives.  We will in what follows
mostly concentrate on a particular one, but the reader in urged to
remember that the formulas above are more adaptable than we shall make
out.  

Let us now turn to the computation of interesting observables, the
most important of which is the total current through the system.  This
is given by the expectation value of the operator (we put the
electron's charge to one)
\begin{eqnarray}
I_T (x) = i \sum_{j \sigma} v^j \left[ \Psi_{L \sigma}^{\dagger} (x)
\Psi_{L \sigma} (x) -
\Psi_{R \sigma}^{\dagger} (x) \Psi_{R \sigma} (x) \right].
\end{eqnarray}
in terms of chiral components.  
Now suppose that for all $ij$, we have $\Delta_L^{ij} = 
|\Delta_L^{ij}|, \Delta_R^{ij} = |\Delta_R^{ij}| e^{-i \chi}$.  
The physical interpretation of this is clear:  the coherence length of
the original superconductor is large enough to subject all channels to
the same pairing phase ({\it i.e.}, we have seen that the bulk pairing
phase gets transferred more or less directly to the phase of the
boundary pairing.  If all the channels are coupled to the
superconductor within a domain of typical size smaller than the
coherence length, then the induced boundary pairing phase is the same
for all channels).  The modulus of the boundary pairings, however, can
still vary from one channel to the other:  microscopic details then
have influence over them.  In 
that case, it is straightforward to show the well-known fact that  
\begin{eqnarray}
I (\chi) = \langle I_T \rangle = -\frac{2}{\beta} \frac{d}{d\chi} \ln Z 
\end{eqnarray}
using sources and an $x$-dependent gauge transformation.  Therefore, a
knowledge of the phase dependence of the partition function allows one
to compute the current without further work, simply by using this
derivative trick. This derivation is valid even in the presence of 
interactions, as long as these can be expressed as current-current 
interactions.  

\begin{figure}
\includegraphics[width=12cm]{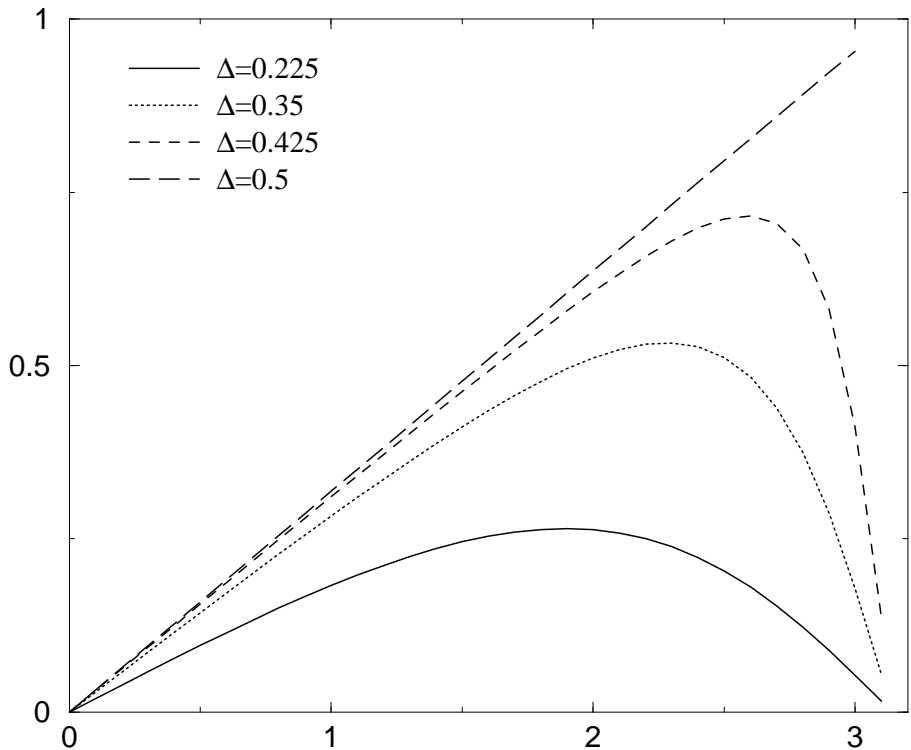}
\caption{The Josephson current at zero temperature for different
values of the boundary parameters as a function of the phase $\chi$.
This is for the 2-channel case, with no interactions, and with equal
pairing amplitudes (that is, all the pairing matrix elements are
equal).  The critical current at zero temperature in this figure and
the one below has been put to 1.} 
\label{Tzerocurrent}
\end{figure}

 At zero temperature, we obtain the simple formula
\begin{eqnarray}
I_0 (\chi) = 2 \frac{d}{d\chi} E_0
\end{eqnarray}
involving the ground-state energy above.  This is plotted in figure
(\ref{Tzerocurrent}) for the case of boundary pairing with equal
entries, i.e. $\Delta^{ij} = \Delta$.  This essentially reproduces the
plot in \cite{AffleckPRB62}, 
done for a single channel, the only difference being that the perfect
current occurs for $\Delta = 1/2$ instead of one.
One can clearly see how the shape of the
current-phase relationship changes from a $\sin \chi$ behaviour for
small boundary pairing (so with normal reflection amplitude $>>$
Andreev reflection amplitude) to the sawtooth function $\chi$ (mod
$2\pi$) for optimal boundary pairing (where Andreev reflection
amplitude $>>$ normal reflection amplitude).  

\begin{figure}
\includegraphics[width=12cm]{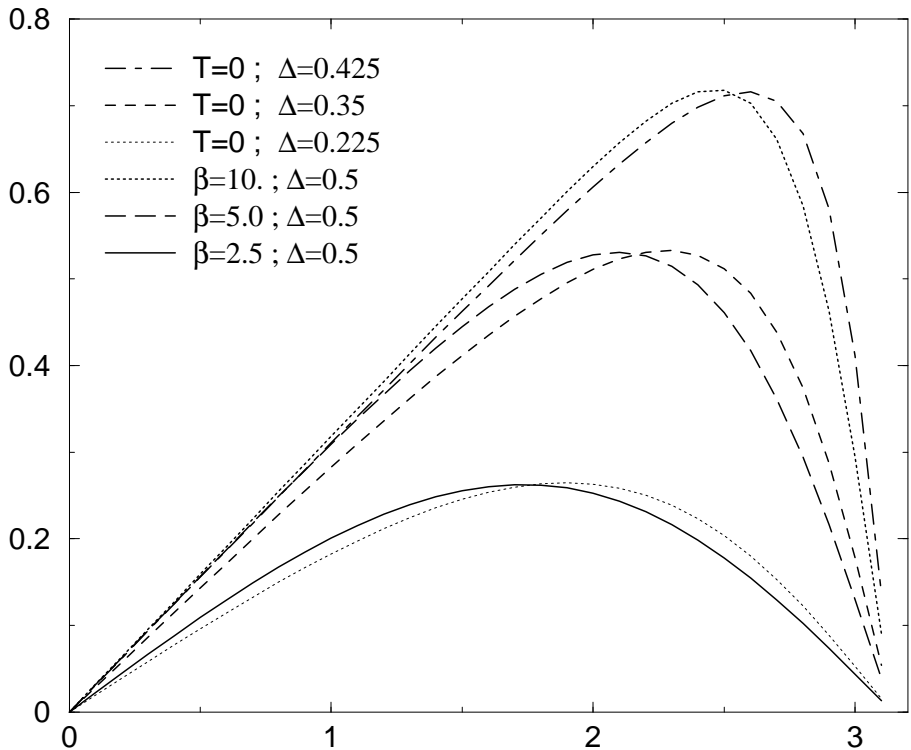}
\caption{The Josephson current at different temperatures for a few 
values of the boundary parameters as a function of the phase $\chi$.}
\label{Tdiffzero}
\end{figure}

At finite temperatures, we can obtain a closed form expression by
computing the $\chi$-derivative by hand.  Defining the functions
\begin{eqnarray}
\tilde{F} (-i\eta) = e^{2i \arctan({\bf v}^{-1} {\bf M}_L)} e^{-i
\sigma_z \otimes {\bf k}_F R - {\bf 1} \otimes {\bf v}^{-1} \eta R} i
\left[ \sigma_z, e^{2i \arctan({\bf v}^{-1} {\bf M}_R)} \right] e^{-i
\sigma_z \otimes {\bf k}_F R - {\bf 1} \otimes {\bf v}^{-1} \eta R}
\label{tcurrent0}
\end{eqnarray}
we can then write
\begin{eqnarray}
I(\chi) = - \beta^{-1} \sum_{n=0}^{\infty} \mbox{Tr} \left[ [1 + F(-i
\eta_n)]^{-1} \tilde{F} (-i\eta_n) + \mbox{h.c.} \right]
\label{tcurrent}
\end{eqnarray}
where $\eta_n = \frac{2\pi}{\beta} (n + 1/2)$. Such a current is plotted in
figure (\ref{Tdiffzero}) for different temperatures and for a few values
of the boundary parameters.

The appearance of the commutator in equation (\ref{tcurrent0}) is
easily interpreted:  the terms in (\ref{bdrymatrix}) that do not
commute with the $\sigma_z$ matrix are precisely the boundary pairing
terms driving Andreev reflection, and therefore the Josephson
current. 

This completes our discussion of the noninteracting case.  From our
formulas, it is easy to compute other transport properties, or to
adapt to different systems with, for example, boundary terms breaking
the spin symmetry.  We leave these for now, and turn instead to the
effects of interactions in the bulk.

\section{The interacting case}

In this section, we study two general classes of interacting one-dimensional 
systems at finite size, coupled to the boundary conditions that simulate the 
superconducting leads to which the real system is coupled.
We consider first the case of a multichannel Luttinger liquid in which the 
different channels only interact through the boundaries.
Then we study the case of a carbon nanotube which can be modelled as a two channel 
Luttinger liquid with a particular choice of interchannel coupling.
In both cases, we consider as the starting point the system coupled to either 
normal or Andreev boundary conditions, and introduce more general boundary 
conditions perturbatively.
These lead to different types of behaviour, which we detail by presenting
computations of the Josephson current (either perturbatively or
exactly depending on the situation), and of the superconducting order parameter
profile in the bulk of the system induced by the presence of the superconducting
leads (the ``proximity effect'').

%%%%%%%%%%%%%%%%%%%%%%%%%%%%%%%%%%%%

\subsection{Multichannel Luttinger liquid}

It has long been appreciated theoretically that the interacting
Luttinger liquid provides a one-dimensional metallic counterpart to the 
higher-dimensional Fermi-liquid state. However, there are no fermionic
quasi-particles in a Luttinger liquid and the elementary excitations are
rather bosonic collective charge and spin fluctuations dispersing with
different velocities. 
One needs to take interactions into account right 
from the beginning by using the bosonization technique. The power of
this technique  
resides in the fact that it allows to express an interacting fermionic
Hamiltonian  
in terms of a noninteracting bosonic one.  Correlation functions of
fermion operators become correlation functions of (combinations of)
bosonic vertex operators, which can be computed exactly.

The bosonization procedure has been extensively reviewed in the
literature \cite{AlexeiBook,bos,senechal}. 
Here we will only discuss in detail the particularities  due to finite 
size and finite temperature, and how they depend on the boundary 
conditions.  We start by noting that the fermion 
operators can be written in terms of bosons as
\begin{eqnarray}
& \; \; \; \Psi_{L \sigma}^j =
\frac{\eta_{L \sigma}^j}{\sqrt{4\pi \delta}} e^{-i \sqrt{2\pi} \left
[ \cosh \xi_c^j \phi_{cL}^j + 
\sinh \xi_c^j \phi_{cR}^j + \sigma [ \cosh \xi_s^j \phi_{sL}^j + \sinh
\xi_s^j \phi_{sR}^j] \right]}, \nonumber \\
&\Psi_{R \sigma}^j =
\frac{\eta_{R \sigma}^j}{\sqrt{4\pi \delta}} e^{i \sqrt{2\pi}
\left[ \sinh \xi_c^j \phi_{cL}^j +
\cosh \xi_c^j \phi_{cR}^j + \sigma [ \sinh \xi_s^j \phi_{sL}^j + \cosh
\xi_s^j \phi_{sR}^j] \right]}
\label{LLBosonization}
\end{eqnarray}
where the $\eta$'s are Klein factors obeying a diagonal Clifford algebra,
ensuring anticommutation of 
different fermionic species.  The index $j=1,...N$ labels the channel, and 
$\sigma = \pm$ labels the spin.  The interaction parameters $\xi_{c,s}^j$
vanish in the noninteracting case, and  are related to other
conventional notations by 
\begin{eqnarray}
e^{2 \xi_{c,s}^j} = K_{c,s}^j.
\end{eqnarray}
Repulsive interactions correspond to taking $\xi < 0 $ $(K < 1)$, whereas
attractive interactions correspond to taking $\xi > 0 $ $(K > 1)$.  
In the following, we shall make use of both of these notations
interchangeably.

The Hamiltonian of the bulk interacting theory is that of free chiral left-
and right-moving bosons for the charge and spin sectors:
\begin{eqnarray}
H = \sum_j \sum_{a = c,s} H^j_a, \hspace{1cm}
H_a^j = \int_0^R dx v_a^j \left[ (\partial_x \phi_{aL}^j)^2 +
(\partial_x \phi_{aR}^j)^2 \right].
\end{eqnarray}
The canonical commutation relations in the interval $]0, R[$ are 
\begin{eqnarray}
\left[ \phi_{a L}^j (x,t), \partial_x \phi_{b L}^{j'} (x',t)
\right] = \frac{i}{2} \delta^{j j'} \delta_{ab} \delta (x - x'),
\nonumber \\
\left[ \phi_{a R}^j (x,t), \partial_x \phi_{b R}^{j'} (x',t)
\right] = -\frac{i}{2} \delta^{j j'} \delta_{ab} \delta (x - x').
\end{eqnarray}

When working in the Hamiltonian formalism, our first objective
is to provide adequate mode expansions for our fields.   
In a finite-size setup such as the one we are considering, the
boundary conditions influence 
the particulars of this mode expansion:  boundary conditions are in
fact constraints on the fields that have to be properly taken into
account.  Unfortunately, in the interacting case, we cannot simply
write down some interpolating boundary conditions like we have done
for the fermionic fields in the noninteracting case.  Instead, one has
to impose some tractable boundary conditions, and use a perturbative
setup around them.  Two cases are detailed
below, which we have already introduced in the noninteracting case:
normal (eq. \ref{normalbcs}) and Andreev (eq. \ref{Andreevbcs}) boundary conditions.

%%%%%%%%%%%%%%%%%%%%%%%%%%%%%%%%%%%%%%%%%%%%%%

\subsubsection{Normal boundary conditions}

The philosophy that we adopt
for now is to start by imposing normal boundary conditions on the
fermions (\ref{normalbcs}).  In terms of the charge and spin bosons,
this means that we fix (being careful with the zero modes)
\begin{eqnarray}
&\phi_c^j (0,t) = \phi_{cL}^j (0,t)+ \phi_{cR}^j (0,t)
=\sqrt{\frac{\pi}{2 K_c^j}} [n_{\uparrow}^j +
n_{\downarrow}^j] \equiv \alpha_c^j, 
\hspace{.8cm} \phi_s^j (0,t) = \phi_{sL}^j (0,t)+ \phi_{sR}^j (0,t) 
= \sqrt{\frac{\pi}{2 K_s^j}}[n_{\uparrow}^j -
n_{\downarrow}^j] \equiv \alpha_s^j, 
\nonumber \\
&\phi_c^j (R,t) = \sqrt{\frac{\pi}{2 K_c^j}} [m_{\uparrow}^j +
m_{\downarrow}^j + 1 - \frac{2 k_F^j R}{\pi}] \equiv \beta_c^j, 
\hspace{.8cm} \phi_s^j (R,t) = \sqrt{\frac{\pi}{2 K_s^j}}[m_{\uparrow}^j -
m_{\downarrow}^j] \equiv \beta_s^j, 
\label{NBCs}
\end{eqnarray}
with $n,m \in \mathbb{Z}$.  
The mode expansions satisfying these boundary conditions can  be
explicitly written down as 
\begin{eqnarray}
\phi_{a L}^j (x,t) = 
\frac{\hat{\phi}_{aL}^{j}}{2} +
\hat{\Pi}_{a}^j 
\frac{x + v^j_a t}{2R} + \zeta_{a L}^j (x,t), \nonumber \\
\phi_{a R}^j (x,t) = 
\frac{\hat{\phi}_{aR}^{j}}{2} +
\hat{\Pi}_{a}^j 
\frac{x - v^j_a t}{2R} + \zeta_{a R}^j (x,t),
\label{LLmeNbc}
\end{eqnarray}
with dynamical parts
\begin{eqnarray}
\zeta_{a L}^j (x,t) = \frac{i}{\sqrt{4\pi}} \sum_{n \in \mathbb{Z}}
\frac{1}{n} b_{a n}^j e^{-i \pi n(x + v_j t)/R} = - \zeta_{a R}^j (-x,
t) 
\end{eqnarray}
and commutation rules
\begin{eqnarray}
[\hat{\phi}^j_{aL}, \hat{\Pi}^k_{a'}] = -[\hat{\phi}^j_{aR},
\hat{\Pi}^k_{a'}]= i \delta^{jk}
\delta_{a a'},
\hspace{1cm} 
[b_{a n}^j, b_{a' m}^k] = n \delta_{n+m, 0} \delta^{jk}
\delta_{a a'}, 
\end{eqnarray}
with the left and right zero modes commuting.  

The eigenvalues of the
zero modes are set by the boundary conditions to relative values given
by the topological number combinations (\ref{NBCs}):
\begin{eqnarray}
\hat{\phi}_{aL}^j + \hat{\phi}_{aR}^j = 2 \alpha_a^j, \hspace{1cm} \Pi_a^j =
\beta_a^j - \alpha_a^j.
\label{pisfisN}
\end{eqnarray}

In terms of operators, the charge and spin Hamiltonians become
\begin{eqnarray}
H_{a}^j = \frac{v_a^j}{2R} \hat{\Pi}_a^{j 2} + \frac{\pi v_a^j}{2R}
\sum_{n \neq 0} b_{a, -n} b_{a,n} = \frac{v_a^j}{2R} \hat{\Pi}_a^{j 2}
+ \frac{\pi v_a^j}{R} \sum_{n > 0} b_{a, -n} b_{a,n} - \frac{\pi
v_a^j}{24 R}
\label{NBCsHamiltonian}
\end{eqnarray}

Correlation functions can be obtained exactly in the finite-size
geometry we are considering, even in the presence of a finite
temperature.  Our particular choice of fermionic boundary conditions
yields a simple variation on the so-called open boundary conditions
used in \cite{MattssonPRB56}, where the computation of the two-point
correlator is made.  In the Appendix \ref{apA}, we have extended this result
to the general case of the $2n$-point correlation function.  One thing
to keep in mind is that we have computed canonical correlators, not
(imaginary) time-ordered ones.  The latter, however, can be obtained
as combinations of the correlators we have derived.  

The first thing that we are interested in computing is the correction
to the partition function coming from the boundary couplings
representing the effect of the superconductors.  We perform this
computation starting once again from a real-time formalism.  There are
some subtleties when deriving a Matsubara-like formalism in the
particular finite-size geometry that we are using:  we devote 
Appendix  \ref{apC} to this issue.

The perturbing Hamiltonian, after making use of the boundary
conditions, becomes
\begin{eqnarray}
H_1 = H_{1L} + H_{1R}, \nonumber \\
H_{1L} = 2 \sum_{i,j=1}^N \Delta_L^{ij} \left[ \Psi_{L \uparrow}^{i
\dagger} (0) \Psi_{L \downarrow}^{j \dagger} (0) + h.c. \right],
\nonumber \\
H_{1R} = 2 \sum_{i,j=1}^N \Delta_R^{ij} \left[ e^{i\chi + i(k_F^i +
k_F^j)R} \Psi_{L \uparrow}^{i
\dagger} (R) \Psi_{L \downarrow}^{j \dagger} (R) + h.c. \right].
\end{eqnarray}
The correction to the partition function comes
from the usual Matsubara term.  The first order vanishes;  after a bit
of algebra, we are
left with the second order $\chi$-dependent contribution
\begin{eqnarray}
<\rho_{LL}^{(2)}>_{\chi} = 16 \cos \chi \sum_{i,j=1}^N \Delta_L^{ij}
\Delta_R^{ij} \Re \int_0^{\beta} d\tau_1 \int_0^{\tau_1} d\tau_2
\left[ (1-\delta_{i,j}) G^{(2)}_i (-i(\tau_1 - \tau_2), R) G^{(2)}_j
(-i(\tau_1 - \tau_2), R) + \right. \nonumber \\
\left. +  \delta_{i,j} G^{(4)}_{i} (-i(\tau_1 -
\tau_2), R) \right]
\end{eqnarray}
where $\Re$ denotes the real part, and the correlation functions can
be computed using the formulas in Appendix \ref{apA}.  After a bit of
algebra, and some simplifications, we obtain for the two-point
contributions the expressions 
\begin{eqnarray}
G^{(2)}_i (-i\tau,R) = {\cal N}_{1}^{-1} \tilde{Q}_i (-i\tau, R)
\prod_{a=c,s} P_{a,i} \left 
[ F_a^i (-i v_a^i \tau + R) \right]^{-\frac{e^{-2\xi_a^i}}{2}},
\end{eqnarray}
where the zero-mode part is
\begin{eqnarray}
\tilde{Q}_i (-i\tau, R) = - e^{i \frac{k_F^i R \tau_c^i \tau}{
\beta}} \frac{ \theta_1 (\tau_c^i [ \frac{\pi \tau}{2\beta} + k_F^i R]
| \tau_c^i) \theta_4 (\tau_s^i \frac{\pi \tau}{2\beta} | \tau_s^i) + 
\theta_4 (\tau_c^i [ \frac{\pi \tau}{2\beta} + k_F^i R]
| \tau_c^i) \theta_1 (\tau_s^i \frac{\pi \tau}{2\beta} | \tau_s^i)}
{\theta_2 (\tau_c^i k_F^i R | \tau_c^i) \theta_3 (0 | \tau_s^i) + 
\theta_3 (\tau_c^i k_F^i R | \tau_c^i) \theta_2 (0 | \tau_s^i)}.
\end{eqnarray}
The reader unfamiliar with $\theta$-functions is referred to Appendix
B.  In order to make the bulk of our paper less clogged up, we have
defined all functions and parameters in Appendix A.  Suffice it to say
here that the dynamical functions $F$ are given by 
\begin{eqnarray}
F_a (z) = \frac{\vartheta_1 (\frac{\pi z}{2R} | \omega_a)}
{\vartheta_1 (-i \frac{\pi \alpha}{2R} | \omega_a)} 
\end{eqnarray}
in which $\alpha$ is an implicit cutoff, i.e. $z \rightarrow z - i 
\alpha$.  The periods of the theta functions are
\begin{eqnarray}
\omega_a = i \frac{v_a \beta}{2R}.
\end{eqnarray}
The phases are $P_{a,i} = e^{i\frac{\pi (1 + e^{-2\xi_a^i})}{4}}$. 
For the four-point function, we get
\begin{eqnarray}
G^{(4)}_i (-i\tau,R) = {\cal N}_{2}^{-1} \tilde{Q}^{(4)}_i P_{c,i}^4
\left[ F_c^i (-i v_c^i \tau + R) \right]^{-2 e^{-2\xi_c^i}}
\end{eqnarray}
where the zero-mode part is
\begin{eqnarray}
\tilde{Q}^{(4)}_i (-i\tau,R) = - e^{i \frac{2 k_F^i R \tau_c^i
\tau}{\beta}} \frac{ \theta_2 (\tau_c^i [ \frac{\pi \tau}{\beta}
+ k_F^i R] 
| \tau_c^i) \theta_3 (0| \tau_s^i) - 
\theta_3 (\tau_c^i [ \frac{\pi \tau}{\beta} + k_F^i R]
| \tau_c^i) \theta_2 (0| \tau_s^i)}
{\theta_2 (\tau_c^i k_F^i R | \tau_c^i) \theta_3 (0 | \tau_s^i) + 
\theta_3 (\tau_c^i k_F^i R | \tau_c^i) \theta_2 (0 | \tau_s^i)}.
\end{eqnarray}
A few comments might be in order here.  First of all, the definitions
of the functions used here are to be found in Appendix A.  For a quick
match with the usual correlation function behaviours, however, one
just has to remember that the functions $F_a (z)$ behave like $\propto
z$ for zero temperature and infinite system size.  Therefore, the
scaling exponent of correlation functions is just (minus) the exponent
of $F$ in the expression encountered.  This gives the usual bulk
power-law correlation functions so typical in this kind of business.
In particular, at the level of the dynamical contributions through the
$F$ functions, the charge and spin sector are completely separated, a
phenomenon which is usually called spin-charge separation.  The two
types of excitations, spin and charge modes, move at different
velocities, seemingly splitting the electron in different parts.
Notice the crucial fact, however, that the zero-mode part above does
not separate at all in distinct charge and spin sectors.  Via the zero
modes, these two sectors remain completely entangled, unless
we neglect all finite-size effects.  In the limit of an infinite size,
the zero-mode parts simplify to trivial phases.  We are, however,
interested in finite-size effects, so we keep these contributions
throughout our treatments.  
This will turn out to be true of all correlators we shall compute in
all cases:  the dynamical sector readily factorizes into charge and
spin sectors, but the zero-modes entangle the two in a complicated
manner.  

The expression for the Josephson current to this order in perturbation
theory is easily obtained by applying the derivative trick:
\begin{eqnarray}
I(\chi) = 32 \sin \chi \sum_{i,j=1}^N \Delta_L^{ij}
\Delta_R^{ij} \Re \frac{1}{\beta} \int_0^{\beta} d\tau_1
\int_0^{\tau_1} d\tau_2 
\left[ (1-\delta_{i,j}) G^{(2)}_i (-i(\tau_1 - \tau_2), R) G^{(2)}_j
(-i(\tau_1 - \tau_2), R) + \right. \nonumber \\
\left. +  \delta_{i,j} G^{(4)}_{i} (-i(\tau_1 -
\tau_2), R) \right].
\label{ILLNBC}
\end{eqnarray}
One thus obtains a $\sin \chi$ behaviour to this order in perturbation
theory, with a normalization that is bilinear in the boundary
pairings, and contains a rather nontrivial function of the system
size, temperature, velocities and interaction parameters.  This
reproduces the expected form of the current near the normal fixed
point.  Higher-order terms in perturbation theory could now be taken
into account, which would start showing deviations from the pure $\sin
\chi$ behaviour.  Moreover, each term in the perturbation theory would
be composed of some nontrivial function of $\chi$ multiplied by a
complicated integral over correlation functions, in other words a
rather daunting function of system size.  These would in turn conspire
to make the current some scaling function of appropriate powers of the
boundary pairings and the system size.  We have not included those
computations here, although they can in principle be performed using
our formulas, given enough patience.

One of the features of equation (\ref{ILLNBC}) is that the current
contains two different types of scaling functions of system size, with
different exponents associated to them.  In the limit of large system
size and zero temperature, we can extract the leading behaviours of
these, and write symbolically 
\begin{eqnarray}
I(\chi) \sim \sum_i I_{\parallel}^i R^{1-2e^{-2\xi_c^i}} + \sum_{ij}
I_{\perp}^{ij} R^{1 - \frac{1}{2} [e^{-2\xi_c^i} + e^{-2\xi_s^i} + i
\rightarrow j]}.
\end{eqnarray}
The first type of scaling comes from Cooper pair tunnelling purely
into one of the channels, whereas the second type comes from Cooper
pair tunnelling into two different channels ({\it i.e.}, the Cooper
pair splits up in two single-electron components penetrating channels
$i$ and $j$).  The particular form of these power laws depends
crucially on the assumption that the system sits close to normal
boundary conditions:  if it is close to Andreev boundary conditions,
the scaling is different, as will be seen later.  One immediate thing
to notice from these scaling functions is the expected fact that the
current gets suppressed by increasing the repulsiveness of the
interactions (that is, making $\xi_c$ smaller), whereas throwing in
more attractive interactions enhances the current.  This is very
natural:  we are dealing with a supercurrent, produced by pairing, and
pairing works hand in hand with attractiveness.
  
Another object of interest is the pairing operator, defined as
\begin{eqnarray}
P^{ij} (x,t) = \Psi_{\uparrow}^i (x,t) \Psi_{\downarrow}^j (x,t) 
\label{opg}
\end{eqnarray}
within the bulk of the system.  The fact that the expectation value of
this operator can be nonvanishing in the bulk of the system represents
the effects of the superconducting boundaries:  these produce the
well-known ``proximity effect'', of which this is one manifestation.
Using the normal boundary conditions,
we can rewrite this pairing 
operator purely in terms of left-movers, and compute its
expectation value to first order in perturbation theory.  The two
boundaries are independent to this order, so we take into account only
the left one for simplicity.  The result is
\begin{eqnarray}
\langle P^{ij} (x,0) \rangle = 4 \Delta_L^{ij} \int_0^{\beta} d\tau
\left[ (1 - \delta_{i,j}) 2 \Re \left\{G_i^{(2)} (-i\tau,x)\right\}
\Re \left\{G_j^{(2)} (-i\tau,x)\right\} - \delta_{ij} [ \Re \left\{ 
G_+^{(4)}(-i\tau, x) \right\} + G_-^{(4)} (-i\tau, x) ] \right].
\label{pairingLLNBC}
\end{eqnarray}
The two-point functions in either channel are here given by slightly
more complicated functions (derived again from the formulas in
Appendix A):  
\begin{eqnarray}
G^{(2)}_i (-i\tau, x) = {\cal N}_{1}^{-1} \tilde{Q}_i (-i\tau, x)
\prod_{a=c,s} \left[ F_a^i (-iv_a^i \tau - x) \right]^{- \frac{1}{4}(1
+ e^{-2\xi_a^i})} \left[ F_a^i (-iv_a^i \tau + x) \right]^{\frac{1}{4}(1
- e^{-2\xi_a^i})} |F_a^i (2x)|^{-\frac{1}{4} \sinh2\xi_a^i}
\end{eqnarray}
where $|F|$ represents the modulus of $F$, and 
the zero mode parts are again given in terms of $\theta$-functions as 
\begin{eqnarray}
\tilde{Q}_i (-i\tau,x) =  e^{i \frac{k_F^i R \tau_c^i \tau}{\beta}}
\frac{ \theta_2 (\tau_c^i [ \frac{\pi \tau}{2\beta} + k_F^i R] 
+ \frac{\pi x}{2R} | \tau_c^i) \theta_3 (\tau_s^i \frac{\pi
\tau}{2\beta} + \frac{\pi x}{2R} | \tau_s^i) +  
\theta_3 (\tau_c^i [ \frac{\pi \tau}{2\beta} + k_F^i R] + \frac{\pi
x}{2R} 
| \tau_c^i) \theta_2 (\tau_s^i \frac{\pi \tau}{2\beta} + \frac{\pi
x}{2R} | \tau_s^i)} 
{\theta_2 (\tau_c^i k_F^i R | \tau_c^i) \theta_3 (0 | \tau_s^i) + 
\theta_3 (\tau_c^i k_F^i R | \tau_c^i) \theta_2 (0 | \tau_s^i)}.
\end{eqnarray}
The four-point functions are of two types.  This comes from the fact
that, when writing the full pairing operator in terms of left-movers
only, one obtains terms like $\Psi_{L \uparrow} (x) \Psi_{L
\downarrow} (x)$ and $\Psi_{L \uparrow} (-x) \Psi_{L
\downarrow} (x)$.  These turn out to have slightly different
$x$-dependencies.  We label the first type $+$, and the second $-$.

The first type of correlator is
\begin{eqnarray}
G_{+;i}^{(4)} (-i\tau,x) = {\cal N}_{2}^{-1} \tilde{Q}_{+;i}^{(4)}
(-i\tau,R) \left[ F_c (-iv_c 
\tau - x) \right]^{-(1 + e^{-2\xi_c})} \left[ F_c(-iv_c \tau + x)
\right]^{1 - e^{-2\xi_c}} |F_c (2x)|^{-\sinh 2\xi_c} 
\end{eqnarray}
with 
\begin{eqnarray}
\tilde{Q}_{+;i}^{(4)} (-i\tau,x) = e^{i \frac{2 k_F^i R \tau_c^i
\tau}{\beta}} \frac{ \theta_2 (\tau_c^i [ \frac{\pi \tau}{\beta}
+ k_F^i R] +\frac{\pi x}{R} 
| \tau_c^i) \theta_3 (0 | \tau_s^i) +
\theta_3 (\tau_c^i [ \frac{\pi \tau}{\beta} + k_F^i R] + \frac{\pi
x}{R} | \tau_c^i) \theta_2 (0 | \tau_s^i)}
{\theta_2 (\tau_c^i k_F^i R | \tau_c^i) \theta_3 (0 | \tau_s^i) + 
\theta_3 (\tau_c^i k_F^i R | \tau_c^i) \theta_2 (0 | \tau_s^i)}.
\end{eqnarray}
The second type is
\begin{eqnarray}
G_{-;i}^{(4)} (-i\tau,x) = \tilde{Q}_{-;i}^{(4)} (-i\tau,R) 
|F_c (-iv_c \tau + x) |^{-2e^{-2\xi_c}} |F_c (2x)|^{\frac{1}{2}
e^{-2\xi_c}} |F_s (2x)|^{-\frac{1}{2}e^{2\xi_s}}
\end{eqnarray}
with 
\begin{eqnarray}
\tilde{Q}_{-;i}^{(4)} (-i\tau,x) = e^{i \frac{2 k_F^i R \tau_c^i
\tau}{\beta}} \frac{ \theta_2 (\tau_c^i [ \frac{\pi \tau}{\beta}
+ k_F^i R]
| \tau_c^i) \theta_3 (\frac{\pi x}{R} | \tau_s^i) +
\theta_3 (\tau_c^i [ \frac{\pi \tau}{\beta} + k_F^i R] | \tau_c^i)
\theta_2 ( \frac{\pi x}{R} | \tau_s^i)} 
{\theta_2 (\tau_c^i k_F^i R | \tau_c^i) \theta_3 (0 | \tau_s^i) + 
\theta_3 (\tau_c^i k_F^i R | \tau_c^i) \theta_2 (0 | \tau_s^i)}.
\end{eqnarray}
One therefore sees that the three terms in (\ref{pairingLLNBC}) have
different scaling behaviour when interactions are added to the bulk of
the system.  In particular, the behaviour of the pairing expectation
value depends on whether we are looking at channel-diagonal or
cross-channel pairing.  The results can be summarized by the scaling
laws 
\begin{eqnarray}
\langle P^{ij} (x) \rangle \sim 
\delta_{ij} P_{\parallel +} x^{1 - \frac{1}{2} [e^{2\xi_c^i} + 3
e^{-2\xi_c^i}]} + \delta_{ij} P_{\parallel -} x^{1 - \frac{1}{2} [ 3
e^{-2\xi_c^i} + e^{2\xi_s^i}]}
+ (1-\delta_{ij}) P_{\perp} x^{1 -
\frac{1}{8} [ e^{2\xi_c^i} + 3 e^{-2\xi_c^i} + e^{2\xi_s^i} + 3
e^{-2\xi_s^i} + i \rightarrow j]} 
\end{eqnarray}
In these, $x$ represents the distance perpendicularly away from the
superconducting boundary.  The cross-channel term has a decay law that
depends on the interaction parameters in both channels, whereas the
channel diagonal term has two different behaviours, with different
exponents depending on the nature of the operator involved.   
These results are collected in simplified form and
discussed further in section \ref{lastsection}.

\subsubsection{Andreev boundary conditions}

The other case that we  consider, which leads to tractable
calculations, requires extremely good coherence between the original
bulk superconductors and the one-dimensional channel.  This does not
mean a perfect contact in the usual sense of the word, however.  We
invite the reader to consult reference \cite{AffleckPRB62}, where
basic considerations on the level of a simple lattice model point to
the fact that there exists an optimization phenomenon, by which it
should be possible to achieve perfect Andreev reflection at the
boundaries by tuning available parameters.  

We therefore want to reproduce the above results in this case. 
Since the specifics of the boundary conditions are crucial to the
physics of the system, physical quantities like the Josephson current
and the superconducting order parameter profile will be different from the previous
results computed around the normal fixed point. 
We therefore start by imposing Andreev boundary conditions on the
fermions, eq. (\ref{Andreevbcs}).  The right-moving fields can then be
described in terms of left-moving ones via
\begin{eqnarray}
\Psi_{L \sigma} (x) = i \epsilon_{\sigma \sigma'} \Psi_{R
\sigma'}^{\dagger} (-x)
\label{perioA}
\end{eqnarray}
with periodicity conditions
\begin{eqnarray}
\Psi_{L \sigma} (x + 2R) = -e^{i\chi} \Psi_{L \sigma} (x).
\end{eqnarray}
In terms of the charge and spin bosons,
this means that we fix
\begin{eqnarray}
{\tilde \phi}_c^j (0,t) = \phi_{cL}^j (0,t)- \phi_{cR}^j (0,t)=\sqrt{\frac{\pi K_c^j}{2 }} [n_{\uparrow}^j +
n_{\downarrow}^j] \equiv {\tilde \alpha}_c^j, 
\hspace{1cm} \phi_s^j (0,t) = \sqrt{\frac{\pi}{2 K_s^j}}[n_{\uparrow}^j -
n_{\downarrow}^j - {\frac 1 2 }] \equiv \alpha_s^j, 
\nonumber \\
{\tilde \phi_c^j} (R,t) = \sqrt{\frac{\pi  K_c^j}{2}} [m_{\uparrow}^j +
m_{\downarrow}^j + 1 - \frac{\chi^j}{\pi}] \equiv {\tilde \beta}_c^j, 
\hspace{1cm} \phi_s^j (R,t) = \sqrt{\frac{\pi}{2 K_s^j}}[m_{\uparrow}^j -
m_{\downarrow}^j + {\frac 1 2 }] \equiv \beta_s^j, 
\end{eqnarray}
with $n,m \in \mathbb{Z}$.  
The mode expansions satisfying these boundary conditions are the same as for normal boundary conditions 
for the spin channel. For the charge channel we have instead
\begin{eqnarray}
\phi_{c L}^j (x,t) = 
\frac{\hat{\phi}_{cL}^{j}}{2} +
\hat{\Pi}_{c}^j 
\frac{x + v^j_c t}{2R} + \zeta_{c L}^j (x,t), \nonumber \\
\phi_{c R}^j (x,t) = 
\frac{\hat{\phi}_{cR}^{j}}{2} +
\hat{\Pi}_{c}^j 
\frac{-x + v^j_c t}{2R} + \zeta_{c R}^j (x,t),
\label{LLmeAbc}
\end{eqnarray}
with dynamical parts
\begin{eqnarray}
\zeta_{c L}^j (x,t) = \frac{i}{\sqrt{4\pi}} \sum_{n \in \mathbb{Z}}
\frac{1}{n} b_{c n}^j e^{-i \pi n(x + v_j t)/R} =  \zeta_{c R}^j (-x,
t) 
\end{eqnarray}
and commutation rules
\begin{eqnarray}
[\hat{\phi}^j_{cL}, \hat{\Pi}^k_{c}] = [\hat{\phi}^j_{cR},
\hat{\Pi}^k_{c}]= i \delta^{jk},
\hspace{1cm} 
[b_{c n}^j, b_{c m}^k] = n \delta_{n+m, 0} \delta^{jk}, 
\end{eqnarray}
with the left and right zero modes, and charge and spin modes commuting.  

The eigenvalues of the
zero modes are the same as before for the spin modes, and for the charge sector
they become
\begin{eqnarray}
\hat{\phi}_{cL}^j - \hat{\phi}_{cR}^j = 2 {\tilde \alpha}_c^j, \hspace{1cm} \Pi_c^j =
{\tilde \beta}_c^j - {\tilde \alpha}_c^j.
\label{pisfisA}
\end{eqnarray}

The fundamental difference between normal and Andreev boundary
conditions is that for the Andreev case, there exists a Josephson
current and a nonzero pairing operator to zeroth order in perturbation
theory.  That is, we can first of all compute the partition function
in the Andreev case, which carries a phase dependence giving nonzero
$I(\chi)$ by using the derivative trick.  

The partition function is given by
$Z= Tr [e^{-\beta \sum_j (H_c^j + H_s^j)}]$ where $H_a^j$ is the charge or spin
 Hamiltonian for a given channel ( eq. \ref{NBCsHamiltonian}). 
Since the Andreev boundary conditions as well as the interactions are channel diagonal, 
 the total partition function is a product of the partition 
function for each channel. Therefore the total current will be  the 
sum of the contributions of each channel. In other words, we need to compute 
simply the partition function for one LL coupled to Andreev boundary conditions.
 This computation was already done by Maslov {\it et al.} in reference
\cite{MaslovPRB53}. 
Here we only  sketch the derivation. Since the Hamiltonian can be written in 
terms of a zero mode part plus a dynamical part that commute with each other, 
the partition function becomes a product $Z= Z_{zm}(\chi) \times Z_{dyn}$ 
where only the zero modes part depends on $\chi$.  Using the explicit 
expressions for  $\Pi_c$ and $\Pi_s$ given in eqs.
(\ref{pisfisN})  and   (\ref{pisfisA}) respectively,
and following the same steps described in Appendix A 
for the computation of the correlation functions,  we obtain
\begin{eqnarray}
Z_{zm}= e^{{\frac{i{\tilde \tau_c}\chi^2}{4\pi}}}
[{{\vartheta_2 ( {\tilde \tau}_c {\frac \chi 2} | {\tilde \tau}_c)
\vartheta_2 (0 | \tau_s) + \vartheta_3 ( {\tilde \tau}_c {\frac \chi 2} | {\tilde \tau}_c)
\vartheta_3 (0 | \tau_s)}}]
\end{eqnarray}
where ${\tilde \tau}_c=i \beta {\frac{v_c e^{2\xi_c}}{R}}$.
We have dropped the channel index for simplicity. 
The current can be readily obtained by taking the derivative with respect to $\chi$ 
of the above expression for each channel, and adding over the channels. 
Since the general form of the current for arbitrary temperature and length, 
involves derivatives of the $\vartheta_{2,3}$ functions, it 
can not be written in a closed form. The asymptotic values of the current for a given 
channel are:
\begin{equation}
I^j(\chi^j) =e \; {\frac { v_c^j e^{2\xi_c^j}}{ R}}  {\frac {  \chi^j}{\pi }}
\end{equation}
at low temperatures  (${\frac {\beta v_c^j}{R}} >> 1$), and 
\begin{equation}
I^j(\chi^j) = 8 e T     
  e^{- \pi{\frac R \beta} \left({\frac {e^{-2\xi_c^j}}{v_c^j}}  + 
 {\frac {e^{2\xi_s^j}}{v_s^j }}   \right) }  \sin(\chi^j)
\end{equation}
at high temperature (${\frac {\beta v_c^j}{R}} << 1$). This expression reflects 
the fact that at high temperatures, repulsive interactions 
($K_c^j = e^{2\xi^j} < 1$) suppress the Josephson current \cite {MaslovPRB53}.

Here as well, the channel diagonal component of the pairing operator 
(eq. (\ref{opg})) can be computed exactly ,
without the need for perturbation theory. The general expression for 
$P^{ii}(x,t)$ can be obtained from the equations in the Appendix A by taking 
the coordinates of both fermions at the same point. Here we only write the 
asymptotic values of the order paramater at high and low temperatures.

At low temperatures we obtain
\begin{equation}
<P^{ii}(x)> \approx e^{i {\frac x R} \chi } \left(\sin {\left(\frac {\pi
x}{R}\right)}\right)^{-{\frac 1 2 } (e^{-2\xi_c} + e^{2\xi_s})}   
\end{equation}
 In the limit of infinite length this correlator behaves like 
$ <P^{ii}(x)> \approx {\frac {1}{x^{ \gamma } }}$ with $\gamma
=(e^{-2\xi_c} + e^{2\xi_s}  )/2$. This result was already derived by Maslov
{\it et al.} in reference \cite{MaslovPRB53}. 

At high temperatures, the order parameter becomes
\begin{equation}
<P^{ii}(x)> \approx  e^{- \pi{\frac x \beta} ({\frac {e^{-2\xi_c}}{v_c }}  + 
 {\frac {e^{2\xi_s}}{v_s }}   ) } 
\end{equation}
We emphasize again that the two results above are completely
nonperturbative, since the partition function and correlators can be
calculated exactly at finite temperature in this finite-size geometry for what we have termed  Andreev boundary conditions (eq.({\ref{Andreevbcs}})).
It is however possible that in the physical system described by the  
multichannel Luttinger liquid , the coupling to the superconductings leads is 
represented by  non diagonal matrices $\Delta_{R,L}$ . This is certainly  the 
case if  the distance between the channels in the Luttinger liquid  is smaller than the 
coherence length in the superconducting lead. In this case, it is 
necessary to take into account the off diagonal contributions of the boundary 
action, ie, those terms corresponding to $\Delta_{R,L}^{ij} \neq 0$ for $i \neq j$.
This can only be done perturbatively, in an analogous way to the one described in the 
previous subsection. We do not include explicity the calculations here, but 
the contributions of these perturbations to the scaling behaviour of the 
superconducting current and the pairing operator are summarized in section \ref{lastsection}. 

%%%%%%%%%%%%%%%%%%%%%%%%%%%%%%%%%%%%%

\subsection{Carbon nanotubes}

Carbon nanotubes (CN), long, thin cylinders of graphite discovered in 1991 by S.
Iijima \cite{Iij91}, are large molecules that are unique for their size, shape, 
and remarkable physical properties. In fact, they can be 
metallic depending on the wrapping of the tube \cite{EggerPRL79,KanePRL79,Jer98}. 
In the following we will be interested in those CN which are metallic and
are known as ``Single walled nanotubes'' (SWNT).
They exhibit unique quantum wire properties that derive from the
tubes' nanometer scale diameters in combination with the special electronic structure
of graphite. Because of the quantization of circumferential modes, the tube's
electronic states do not form one wide electronic energy band but instead split
into one-dimensional subbands with band onsets at different energies. These
subbands are widely separated in energy, on the scale of 1 eV (much larger than the
room-temperature thermal energy $k_B T$, about 0.025 eV). Only two of the
one-dimensional subbands cross the Fermi energy in metallic carbon nanotubes; all
the current through such tubes is therefore predicted to be carried by only this
pair of subbands. Thus, a single wall carbon nanotube can be described by a pair of 
Luttinger liquids whose velocities are equal to each other, and that interact through
Coulomb interaction.
We shall therefore consider only two conducting spinful channels, i.e. 
(taking spin and chirality into account) eight bosonic fields
$\phi_{a \sigma}^s$ with $ a = L,R$,  $\sigma = \pm$ and $ s = 1,2$.
The main difference between the effective description of a carbon nanotube and the 
previous case of multichannel LL, is that for CN the bulk interactions are not 
channel 
diagonal. This is due to the fact that the Coulomb interaction couples only to the 
total charge density \cite{EggerPRL79,KanePRL79} while the
spin sector remains unaffected by the interactions.  Therefore, we
move to a basis of total and relative bosons:
\begin{eqnarray}
\label{totrelbos}
\phi_{\mu a}^{\tau} = \frac{1}{\sqrt{2}} (\phi_{\mu a}^1 + \tau
\phi_{\mu a}^2).
\end{eqnarray}
with $\tau = \pm, \mu=c,s$, and $a=L,R$. In terms of these operators,
the Hamiltonian becomes (after taking the interactions into account
nonperturbatively)  that of a set of free bosons, 
\begin{eqnarray}
H = \sum_{\mu = c,s} \sum_{\tau = \pm} \int_0^R dx v_{\mu}
\left[ (\partial_x \phi_{\mu L}^{\tau})^2 + (\partial_x \phi_{\mu
R}^{\tau})^2 \right]
\end{eqnarray}
which we can now use to represent our fermions with the bosonization formulas
\begin{eqnarray}
& \; \; \; \Psi_{L \sigma}^{1,2} = \frac{1}{\sqrt{4\pi \delta}} \eta_{L \sigma}^s
e^{-i\sqrt{\pi} [\cosh \xi_c \phi_{c L}^+ + \sinh \xi_c \phi_{cR}^+ 
\pm \phi_{cL}^- + \sigma (\phi_{sL}^+  \pm \phi_{sL}^-)] }, \nonumber \\
&\Psi_{R \sigma}^{1,2} = \frac{1}{\sqrt{4\pi \delta}} \eta_{R \sigma}^s
e^{i\sqrt{\pi} [\sinh \xi_c \phi_{c L}^+ + \cosh \xi_c \phi_{cR}^+ 
\pm \phi_{cR}^- + \sigma (\phi_{sR}^+  \pm \phi_{sR}^-)] }.
\end{eqnarray}
In the above, $\xi_c$ is the interaction parameter representing the
Coulomb interaction that involves the total charge of the system (in
both channels of the nanotube).  

%%%%%%%%%%%%%%%%%%%%%%%%%%%%%%%%%%%%%

\subsubsection{Normal boundary conditions}

As was the case with the Luttinger liquid, in order to set up our
computations correctly in the finite size geometry that we are
considering, we have to specify tractable boundary conditions.  We
start with normal boundary conditions for the fermions, which
correspond in the bosonic language to taking
\begin{eqnarray}
\phi_c^{\pm} (0) = {\frac 1 2 }
 \sqrt{\frac{\pi}{ K_c^\pm}} [n_{\uparrow}^\pm +
n_{\downarrow}^\pm] \equiv \alpha_c^{\pm}, 
\hspace{1cm} \phi_s^{\pm} (0) = \frac{\sqrt{\pi}}{2} 
[n_{\uparrow}^\pm -
n_{\downarrow}^\pm] \equiv \alpha_s^{\pm}, \nonumber \\
\phi_c^{\pm} (R) = {\frac 1 2} 
\sqrt{\frac{\pi}{ K_c^\pm}} [m_{\uparrow}^\pm +
m_{\downarrow}^\pm + 1 \pm 1 - \frac{2 k_F^{\pm} R}{\pi}] \equiv
\beta_c^{\pm},  
\hspace{1cm} \phi_s^{\pm} (R) = \frac{\sqrt{\pi}}{2} [m_{\uparrow}^\pm -
m_{\downarrow}^\pm] \equiv \beta_s^{\pm}.
\end{eqnarray}
where $K_c^+ = K_c$, 
$K_c^-=1$, $n_\sigma ^\pm = n_\sigma^1 \pm n_\sigma ^2$ with
$n_{\sigma}^i$ independent integers, and
$k_F^{\pm} = k_F^1 \pm k_F^2$.
The mode expansions for charge and spin operators coincide with the ones given 
in eq. (\ref{LLmeNbc}).

The second-order correction to the partition function can be
calculated in a similar manner as for the multichannel Luttinger
liquid.  After a bit of algebra, and again making use of the
correlators in the Appendix A, we find the $\chi$-dependent contribution
\begin{eqnarray}
\langle \rho_1^{(2)} \rangle_{\chi} = 16 \cos \chi \sum_{i,j=1,2}
\Delta_L^{ij} \Delta_R^{ij} \int_0^{\beta} d\tau_1 \int_0^{\tau_1}
d\tau_2 \Re \left[ (1-\delta_{ij}) G^{(4)}_{\perp} (-i(\tau_1
-\tau_2), R) + \delta_{ij} G^{(4)}_{\parallel} (-i(\tau_1 -\tau_2), R)
\right], 
\end{eqnarray}
with correlators 
\begin{eqnarray}
G^{(4)}_{\parallel} (-i\tau, R) =  P_+ Q (\mathbf{u}_{\perp})
D_{\parallel} (-i\tau, R), 
\nonumber \\
G^{(4)}_{\perp} (-i\tau, R) = P_+ Q (\mathbf{u}_{\parallel})
D_{\perp} (-i\tau, R), 
\end{eqnarray}
where we have defined $P_+ = e^{i\frac{\pi}{2}[1 + e^{-2\xi_c}]}$ (we
remind the reader that $K_c= e^{2\xi_c}$ is the only nonvanishing
$\xi$ for the nanotubes).  
The dynamical parts are for both correlators equal to
\begin{eqnarray}
D_{\parallel} (-i\tau, R) &=& \left[ F_c
(-i\tau v_c + R) \right]^{-e^{-2\xi_c}} \left[ F_c (-i\tau v_c + R)
\right]^{-1}, \nonumber \\
D_{\perp} (-i\tau, R) &=& \left[ F_c
(-i\tau v_c + R) \right]^{-e^{-2\xi_c}} \left[ F_s (-i\tau v_s + R)
\right]^{-1}, 
\end{eqnarray}
independently of the channel index, 
and the zero-mode part is given in terms of the four-dimensional
$\theta$-function (defined in the second appendix)
\begin{eqnarray}
Q (\mathbf{u}) = e^{i (2k_F R - \pi) \tau_c \tau/\beta}
\frac{\theta (\mathbf{u} + \frac{\tau_c}{4}(2k_F R - \pi) \hat{\mathbf{m}}|
\mathbf{\Omega})}{\theta (\frac{\tau_c}{4}(2k_F R - \pi) \hat{\mathbf{m}}|
\mathbf{\Omega})}
\label{Q}
\end{eqnarray}
with vectors
\begin{eqnarray}
\mathbf{u}_{\perp} = \frac{\pi}{4R} \left( 
\begin{array}{c}
i \tau [e^{-2\xi_c} v_c + v_s] + 2R \\
i \tau [e^{-2\xi_c} v_c - v_s] \\
i \tau [e^{-2\xi_c} v_c - v_s] \\
i \tau [e^{-2\xi_c} v_c + v_s] + 2R
\end{array} \right) \hspace{2cm}
\mathbf{u}_{\parallel} = \frac{\pi}{4R} \left( 
\begin{array}{c}
i \tau v_c [e^{-2\xi_c} + 1] + 2R \\
i \tau v_c [e^{-2\xi_c} + 1] + 2R \\
i \tau v_c [e^{-2\xi_c} - 1] \\
i \tau v_c [e^{-2\xi_c} - 1]
\end{array} \right).
\end{eqnarray}
and  $\hat{\mathbf{m}} =(1,1,1,1)$.
The interaction-dependent period matrix $\mathbf{\Omega}$ is defined
in equation (\ref{Omega}). 
We also remind the reader that $\tau_c = i\frac{\beta v_c
e^{-2\xi_c}}{R}$. To obtain the above results we have used that 
$G^{(4) 12}_{\parallel,\perp}= G^{(4) 21}_{\parallel,\perp}$ due to the symmetry properties of the matrix $\mathbf{\Omega}$.

For the Josephson current, this then implies that, to second-order in
perturbation theory, we find the expected $\sin \chi$ behaviour.  More
precisely, we get
\begin{eqnarray}
I(\chi) = F(\beta, R) \sin \chi, 
\end{eqnarray}
with a normalization amplitude function that depends on all the
parameters of the system, and is given in terms of the integral expression
\begin{eqnarray}
F(\beta,R) = 32 \sum_{i,j} \Delta_L^{ij} \Delta_R^{ij} 
\frac{1}{\beta} \int_0^{\beta} \int_0^{\tau_1} \Re \left
[ (1-\delta_{ij}) G^{(4)}_{\perp} (-i(\tau_1 
-\tau_2), R) + \delta_{ij} G^{(4)}_{\parallel} (-i(\tau_1 -\tau_2), R)
\right].
\end{eqnarray}
As was the case for the multichannel Luttinger liquid, the scaling
behaviour of the current splits up into two different types, depending
on the way the Cooper pair penetrates into the conducting channel.
Here, for the case of normal boundary conditions, the scaling
behaviour is the same for channel-diagonal and cross-channel pair
penetration ({\it i.e.}, doesn't depend on whether the pair penetrates
into one channel only, or splits between the two):
\begin{eqnarray}
I(\chi) \sim (I_{\parallel} + I_{\perp}) R^{-e^{-2\xi_c}}.
\end{eqnarray}
The physical interpretation of this is easy:  as Coulomb interactions
involve the total charge density in both channels together, it doesn't
matter how the pair tunnels in, as the interaction cost is the same in
both cases.

For the pairing operator defined in equation (\ref{opg}), a
first-order perturbative computation similar to that in the Luttinger
liquid case yields
\begin{eqnarray}
\langle P^{ij} (x,0) \rangle = -4 \Delta_L^{ij} \int_0^{\beta} d\tau
\Re \left[ (1-\delta_{ij}) \left( G^{(4)}_{\perp +}(-i\tau,x) +
G^{(4)}_{\perp -} (-i\tau,x) \right) +
\delta_{ij} \left(G^{(4)}_{\parallel +}(-i\tau,x) + G^{(4)}_{\parallel
-}(-i\tau,x) \right) \right],
\end{eqnarray}
with correlators splitting up into two different classes
(channel-diagonal or cross-channel) and again two subclasses ($\pm$,
depending on the particular operator combination;  we refer the reader
to the note at the end of the computation of the pairing for the
multichannel Luttinger liquid with normal boundary conditions):
\begin{eqnarray}
G^{(4)}_{\perp \pm} (-i\tau,x) = Q (\mathbf{u}_{\perp \pm})
e^{i\frac{(1 \pm 1)\pi x}{2 R}} F_{\perp
\pm} (-i\tau, x), \nonumber \\
G^{(4)}_{\parallel \pm} (-i\tau,x) = Q (\mathbf{u}_{\parallel \pm})
e^{i\frac{(1 \pm 1)\pi x}{2 R}} F_{\parallel \pm} (-i\tau, x),
\end{eqnarray}
with dynamical parts given explicitly in terms of powers of
$F$-functions (note in particular that if charge and spin velocity
coincide, then the subsets $\pm$ become identical) 
\begin{eqnarray}
F_{\perp +} (-i\tau, x) &=& \left[ F_c (-i\tau v_c - x)
\right]^{-\frac{1}{2}(1 + e^{-2\xi_c})} \left[ F_c (-i\tau v_c + x)
\right]^{\frac{1}{2} (1 - e^{-2\xi_c})} 
\frac{| F_c (2x) |^{-\frac{\sinh 2 \xi_c}{2}}}{\left[ F_s (-i\tau v_s - x)
\right]}, \nonumber \\
F_{\perp -} (-i\tau, x) &=& | F_c (-i\tau v_c + x) |^{-e^{-2\xi_c}} |
F_s (-i\tau v_s + x) |^{-1} |F_c
(2x)|^{\frac{e^{-2\xi_c}-1}{4}}, \nonumber
\\
F_{\parallel +} (-i\tau, x) &=& \left[ F_c (-i\tau v_c - x)
\right]^{-\frac{1}{2}(1 + e^{-2\xi_c})} \left[ F_c (-i\tau v_c + x)
\right]^{\frac{1}{2} (1 - e^{-2\xi_c})} 
\frac{| F_c (2x) |^{-\frac{\sinh 2 \xi_c}{2}}}{\left[ F_c (-i\tau v_c - x)
\right]}, \nonumber \\
F_{\parallel -} (-i\tau, x) &=& | F_c (-i\tau v_c + x) |^{-e^{-2\xi_c}} |
F_c (-i\tau v_c + x) |^{-1} |F_c(2x)|^{\frac{e^{-2\xi_c}-1}{4}}.
\end{eqnarray}
The zero-mode parts are here given by 
equation (\ref{Q}), and the distance vectors appearing in the
multivariable $\theta$-functions representing the contributions of the
zero modes are defined by
\begin{eqnarray}
\mathbf{u}_{\perp \pm} = \frac{\pi}{4R} \left( 
\begin{array}{c}
i \tau [e^{-2\xi_c} v_c + v_s] \pm 2x \\
i \tau [e^{-2\xi_c} v_c - v_s] \\
i \tau [e^{-2\xi_c} v_c - v_s] \\
i \tau [e^{-2\xi_c} v_c + v_s] + 2x
\end{array} \right) \hspace{2cm}
\mathbf{u}_{\parallel \pm} = \frac{\pi}{4R} \left( 
\begin{array}{c}
i \tau v_c [1 + e^{-2\xi_c}] \pm 2x \\
i \tau v_c [1 + e^{-2\xi_c}] + 2x \\
i \tau v_c [-1 + e^{-2\xi_c}] \\
i \tau v_c [-1 + e^{-2\xi_c}]
\end{array} \right).
\end{eqnarray}
We therefore again find three different scaling behaviours for the
pairing operator, as in the multichannel Luttinger liquid case.  The
scaling of the pairing operator expectation value again splits up in
two different types of behaviour, which we can summarize as
\begin{eqnarray}
\langle P^{ij} (x) \rangle \sim ((1-\delta_{ij}) P_{\perp +} +
\delta_{ij} P_{\parallel +}) x^{-\frac{1}{4}[e^{2\xi_c} + 3
e^{-2\xi_c}]} +
((1-\delta_{ij}) P_{\perp -} +
\delta_{ij} P_{\parallel -}) x^{-\frac{1}{4}[1 + 3
e^{-2\xi_c}]}.
\end{eqnarray}
First of all, we note again that the scaling of the $+$ operators
coincide, as does the scaling of the $-$ operators.  That is, it's not
important (as far as scaling is concerned) how the Cooper pair
penetrates the nanotube.  These
results are summarized and discussed in section \ref{lastsection}.

%%%%%%%%%%%%%%%%%%%%%%%%%%%%%%%%%%%%%

\subsubsection{Andreev boundary conditions}

For the relative bosons suitable to describe carbon nanotubes (\ref{totrelbos}), 
Andreev boundary conditions can be written as
\begin{eqnarray}
\tilde{\phi}_c^{\pm} (0) = \frac{\sqrt{\pi K_c^{\pm}}}{2}
[n_{\uparrow}^\pm +
n_{\downarrow}^\pm] \equiv \alpha_c^{\pm}, 
\hspace{1cm} \phi_s^{\pm} (0) = \frac{\sqrt{\pi}}{2} [n_{\uparrow}^\pm -
n_{\downarrow}^\pm- {\frac  1 2} \mp {\frac 1 2} ] \equiv \alpha_s^{\pm}, \nonumber \\
\tilde{\phi}_c^{\pm} (R) = \frac{\sqrt{\pi K_c^\pm}}{2}
[m_{\uparrow}^\pm +
m_{\downarrow}^\pm - 1 \mp 1 - \frac{\chi^{\pm}}{\pi}] \equiv
\beta_c^{\pm},  
\hspace{1cm} \phi_s^{\pm} (R) = \frac{\sqrt{\pi}}{2} [m_{\uparrow}^\pm -
m_{\downarrow}^\pm +{\frac 1 2} \pm {\frac 1 2}] \equiv \beta_s^{\pm},
\end{eqnarray}
where $n_\sigma ^\pm = n_\sigma^1 \pm n_\sigma ^2$ with
$n_{\sigma}^i$ independent integers, and $\chi^{\pm} = \chi^1 \pm
\chi^2$. The mode expansion for charge and spin operators coincide
with the one given in eqs. (\ref{LLmeAbc}) and (\ref{LLmeNbc})
respectively. 

In order to compute the superconducting current, we  consider the partition
 function. Once again, the only part of the partition function which is 
$\chi$ dependent is the zero mode part $Z_{zm}$. Since the two channels 
constituting the nanotube are coupled to the superconductor leads at the 
boundary in the same way, the two phases must  satisfy $\chi^1 = \chi^2=\chi$, and 
the current is determined by  
$ Z_{zm} (\chi)$
\begin{eqnarray}
Z_{zm}  (\chi)= e^{{\frac{i{\tilde \tau_c}\chi^2}{2\pi}}}
\theta ({\frac{{\tilde \tau_c}\chi}{4}} \hat{\mathbf{m}} |
\tilde{\mathbf{\Omega}})  
\end{eqnarray}
where ${\tilde \tau}_c = i \beta {\frac{v_c e^{2\xi_c}}{R}}$, and
$\hat{\mathbf{m}} =(1,1,1,1)$.  
The multivariable 
theta function $\theta$ and the matrix period $\tilde{\mathbf{\Omega}}$, 
defined in  Appendix B and A respectively, appear naturally when computing correlation 
functions for carbon nanotubes, where the electronic interactions are not channel 
diagonal.

The total current is obtained by deriving $\ln Z_{zm}$ 
with respect to 
the phase  $\chi$. Its asymptotic values are in this case given by the expressions
\begin{equation}
I(\chi) =e \; {\frac {2 e^{2\xi_c} v_c }{ R}} {\frac {  \chi}{\pi }} 
\end{equation}
at low temperature  (${\frac {\beta v_c^j}{R}} >> 1$), and 
\begin{equation}
I(\chi) = 16 e T     
  e^{- \pi{\frac R \beta} ({\frac {e^{-\xi_c} \cosh{\xi_c}}{v_c }}  +  
{\frac {1}{v_s }}   ) }  \sin \chi 
\end{equation}
at high temperature (${\frac {\beta v_c^j}{R}} << 1$). 

A physical realization of the coupling between a nanotube and a
superconductor would however produce boundary pairing amplitudes
$\Delta^{ij}$ which should be approximately equal for all the channels
(that is, even for the off-diagonal components).  This is a
fundamental difference between the present case and what one could
expect from a two-channel Luttinger liquid:  in the latter case, the
off-diagonal pairing components could be taken to vanish if the
contacts between the original superconductor and the two Luttinger
liquids were separated by a finite distance.
The diagonal components of the boundary pairing are easy to treat
nonperturbatively as seen above.  In view of the last comments,
however, we also want to include another contribution to the partition
function coming from the channel off-diagonal boundary pairings. The perturbing Hamiltonian 
is given by

\begin{eqnarray}
H_1 = H_{1L} + H_{1R}, 
\end{eqnarray}
where
\begin{eqnarray}
H_{1L} = {\frac 1 2} \left[ \Delta_L^{12}  \Psi_{ \uparrow}^{1
\dagger} (0) \Psi_{ \downarrow}^{2 \dagger} (0) +
 \Delta_L^{21}  \Psi_{ \uparrow}^{2
\dagger} (0) \Psi_{ \downarrow}^{1 \dagger} (0) \right] + h.c.,
\nonumber \\
H_{1R} = {\frac 1 2 } \left[ \Delta_R^{12}  \Psi_{ \uparrow}^{1
\dagger} (R) \Psi_{L \downarrow}^{2 \dagger} (R) 
+\Delta_R^{21}  \Psi_{ \uparrow}^{2
\dagger} (R) \Psi_{L \downarrow}^{1 \dagger} (R) \right] + h.c..
\label{PertHamNTABC}
\end{eqnarray}

The first non vanishing correction is the second order one, which is given by
\begin{eqnarray}
<\rho_{CN}^{(2)}>_{\chi} = 2
 \Delta_L
\Delta_R \Re \int_0^{\beta} d\tau_1 \int_0^{\tau_1} d\tau
e^{i2k_F R -i {\tilde \tau_c}{\frac \chi \beta}\tau}
\left(  {\frac {\theta ({\mathbf u}_{\uparrow}+{\frac{{\tilde \tau_c}\chi}{4}} \hat{\mathbf{m}} |
\tilde{\mathbf{\Omega}})  +
 \theta ({\mathbf u}_{\downarrow}+{\frac{{\tilde \tau_c}\chi}{4}} \hat{\mathbf{m}} |
\tilde{\mathbf{\Omega}})}{\theta ({\frac{{\tilde \tau_c}\chi}{4}} \hat{\mathbf{m}} |
\tilde{\mathbf{\Omega}})}}   
\right) \nonumber \\
\left[ F_c (-iv_c \tau  +R) \right]^{-{\frac { e^{2\xi_c}+1}{2}}} 
\left[ F_c (-iv_c \tau  -R) \right]^{-{\frac { e^{2\xi_c}-1}{2}}} 
\left[ F_s (-iv_s \tau +R) \right]^{-1}
\end{eqnarray}
where $\Re$ denotes the real part, the vectors ${\mathbf{u}_\uparrow}$ and 
${\mathbf{u}_\downarrow}$ are defined by
\begin{eqnarray}
{\mathbf{u}_\uparrow} = \frac{\pi}{4R} \left( 
\begin{array}{c}
-i (-v_c e^{2\xi_c} +v_s)\tau \\
-2R +i (v_c e^{2\xi_c} +v_s)\tau\\
-2R +i (v_c e^{2\xi_c} +v_s)\tau \\
-i (-v_c e^{2\xi_c} +v_s)\tau
\end{array} \right), \hspace{2cm}
{\mathbf{u}_\downarrow} = \frac{\pi}{4R} \left( 
\begin{array}{c}
-2R +i (v_c e^{2\xi_c} +v_s)\tau\\
-i (-v_c e^{2\xi_c} +v_s)\tau \\
-i (-v_c e^{2\xi_c} +v_s)\tau \\
-2R +i (v_c e^{2\xi_c} +v_s)\tau
\end{array} \right).
\end{eqnarray}
respectively, and the correlation functions were computed using the formulas 
in Appendix \ref{apA}.
The correction to the superconducting current will be given by 
$I^{(2)} (\chi)= -{\frac 2 \beta}
{\frac {\partial <\rho_{CN}^{(2)}>_{\chi} }{\partial \chi}}$. 
Even though we can not give a closed 
expression for this correction, we can study  some simple limits.
In particular, we can show that in the limit of low temperatures 
$I^{(2)} (\chi)$ will depend on the length of the system as $ {\frac
{1}{R^{K_c}}}$ times a complicated function of  
$\chi$, representing deviations from the sawtooth form of the perfect
current. Thus, the off diagonal contribution to the superconducting
current  
introduces a dependence on the length of the system which is
interaction dependent. Due to the general structure of the correlation
functions given in Appendix \ref{apA} , it is simple to see that
higher order terms in the perturbative expansion will contribute with
higher order powers of $ {\frac {1}{R^{K_c}}}$ to the superconducting
current at zero temperature.

The order parameter can be computed exactly if we only consider channel 
diagonal Andreev boundary conditions. In this case, using the boundary
 conditions (eq. \ref{perioA}) we can express the pairing  operator  
(eq. (\ref{opg})) as
\begin{eqnarray}
<P^{ii}(x) > = -i \left[ 
<\Psi_{L\uparrow}^{\dagger i} (-x,t) \Psi_{L\uparrow}^i (x,t) > +
<\Psi_{L\downarrow}^{\dagger i} (-x,t) \Psi_{L\downarrow}^i (x,t) >
 \right]   \nonumber \\
= -i 
\left(  {\frac {\theta ({\mathbf u}_{\uparrow}^i+{\frac{{\tilde
\tau_c}\chi}{4}} \hat{\mathbf{m}} | 
\tilde{\mathbf{\Omega}})  +
 \theta ({\mathbf u}_{\downarrow}^i+{\frac{{\tilde \tau_c}\chi}{4}}
\hat{\mathbf{m}} | 
\tilde{\mathbf{\Omega}})}{\theta ({\frac{{\tilde \tau_c}\chi}{4}} \hat{\mathbf{m}} |
\tilde{\mathbf{\Omega}})}}   
\right) \nonumber \\
\left[ F_c (-2x) \right]^{-{\frac { e^{-2\xi_c}+3}{8}}} 
\left[ F_c (2x) \right]^{-{\frac { e^{-2\xi_c}-1}{8}}} 
\left[ F_s (-2x) \right]^{-{\frac 1 2 }}
\end{eqnarray}
where ${\mathbf u}_\sigma^1 = {\frac {\pi x}{2R}} (1+\sigma,
1-\sigma,0,0)$ and  ${\mathbf u}_\sigma^2 = {\frac {\pi x}{2R}}
(0,0,1+\sigma, 1-\sigma)$. 
The  contributions to $<P^{ii}(x) >$ of the form $<\Psi_{L\downarrow}^{i} (x,t) \Psi_{L\uparrow}^i (x,t) >$  as well as the  off-diagonal components of 
the pairing operator ($<P^{12}(x) >,<P^{21}(x) >$) 
vanish unless we take into account off diagonal terms in the boundary 
couplings.  

At low temperatures we obtain
\begin{equation}
<P(x)^{ii}> \approx e^{i {\frac x R} \chi } (\sin {\frac {\pi x}{R}})^{-{\frac 1 4 } (e^{-2\xi_c}+3)}  
\end{equation}
 In the limit of infinite length  this correlator behaves like 
$< P(x)^{ii}> \approx {\frac {1}{x^{ \gamma_\parallel } }}$ with $\gamma_\parallel =( e^{-2\xi_c}+ 3 )/4$. 

At high temperatures, the order parameter becomes
\begin{equation}
<P(x)^{ii}> \approx  e^{- \pi{\frac {x} {2\beta}} ({\frac {e^{-2\xi_c}+1}{v_c  }}  + 
 {\frac {2}{v_s }}   ) } 
\end{equation}

 However, this is not the physical 
case since the LL constituting the CN couple to the boundary in the 
same way, introducing Andreev reflections between the two channels. Thus, 
we need to include perturbative corrections around 
the channel diagonal case. The first order correction due to the 
off-diagonal 
Andreev terms (eqs. (\ref{PertHamNTABC}))  vanish for the diagonal 
components of the  pairing 
operator $<P^{ii}>$, but give 
a finite contribution for $<P^{12}>$, and $<P^{21}>$.  Since to this order the 
two boundaries remain uncoupled, for simplicity  we only  give the correction 
due to the left boundary
\begin{eqnarray}
\langle P^{12} (x,0) \rangle ^{(1)} = - i \Delta_L \int_0^{\beta} d\tau
\left[e^{i2k_F R} G^{12 (4)}_{\downarrow +}(-x, i\tau) +
e^{-i2k_F R} G^{12 \; (4)}_{\uparrow +}(-x, -i\tau) \right]
\end{eqnarray}
with correlators
\begin{eqnarray}
G^{12 (4)}_{\sigma +}(-x,i\tau) = e^{i{\frac \chi R}(x+v_c e^{2\xi_c} \tau)} 
  {\frac {\theta ({\mathbf u}_{\sigma}+{\frac{{\tilde \tau_c}\chi}{4}} 
\hat{\mathbf{m}} |\tilde{\mathbf{\Omega}}) }
{\theta ({\frac{{\tilde \tau_c}\chi}{4}} \hat{\mathbf{m}} |\tilde{\mathbf{\Omega}})}} 
\left[ F_c (i \tau v_c -x) \right]^{-{\frac {e^{2\xi_c} +1}{2}}}
\left[ F_c (i \tau v_c  +x) \right]^{-{\frac {e^{2\xi_c} -1}{2}}} \nonumber \\
|F_c(2x)|^{\frac {\sinh 2\xi_c}{2}}
\left[ F_s (i \tau v_s -x) \right]^{-1}
\end{eqnarray}
where
the distance vectors
are defined by
\begin{eqnarray}
\mathbf{u}_{\downarrow} = \frac{\pi}{4R} \left( 
\begin{array}{c}
-i \tau [e^{2\xi_c} v_c - v_s]  \\
-i \tau [e^{2\xi_c} v_c + v_s] +2x \\
-i \tau [e^{2\xi_c} v_c + v_s] +2x \\
-i \tau [e^{2\xi_c} v_c - v_s] 
\end{array} \right) \hspace{2cm}
\mathbf{u}_{\uparrow} = \frac{\pi}{4R} \left( 
\begin{array}{c}
i \tau [e^{2\xi_c} v_c + v_s]  +2x \\
i \tau [e^{2\xi_c} v_c - v_s]  \\
i \tau [e^{2\xi_c} v_c - v_s] \\
i \tau [e^{2\xi_c} v_c + v_s] +2x
\end{array} \right).
\end{eqnarray}
Thus, there is only one type of non vanishing contributions to this 
operator (to first order in perturbation theory), namely, what we have called
$G^{(4)}_+$. The other terms present in the pairing operator of the form 
$(\Psi_{L\uparrow}^{ i} (-x,t) \Psi_{L\uparrow}^{\dagger i} (x,t)) $
do not contribute to the expectation value of the off diagonal pairing
operator to this order,  due to  
the particular form of the boundary perturbation.
 
In general, this expectation value will be a complicated function of
the distance to the superconductor $x$. However, in the limit of zero
temperature, and for very large systems, we can show that  
$\langle P^{12} (x,0) \rangle ^{(1)}  \approx {\frac {1}{x^
{ \gamma_\perp } }}$ with $\gamma_\perp =(e^{-2\xi_c} + 3e^{2\xi_c}
)/4$.

\section{Summary and discussion}
\label{lastsection}

Up to this point,  we have presented most computations
without making any approximations with respect to system size or temperature. 
This results in rather complex expressions for correlators and their related 
physical quantities which are challenging to interpret straightforwardly.  
We have done this because it is feasible, and represents a very good starting
point for seeking a more precise exact solution to the type of problem
that we have considered.  For example, the exact correlators that we
have computed would have to be reproduced by any other method claiming
to be capable of dealing with similar systems (we are thinking here of
for example massless limits of form factors for the sine-Gordon
model).  

In order to make the important results more accessible, and to paint a
more transparent picture of the systems we considered, we provide here a 
short summary of our results.  We remind
the reader that the two systems that we have dealt with, namely
multichannel Luttinger liquids and carbon nanotubes, have different
bulk interaction structures.  We have considered charge and spin
interactions which are channel-diagonal for the Luttinger liquids,
whereas the interaction for the nanotubes is of the form of a Coulomb
term involving the total charge of the system.  This has an influence
on the specifics of the bosonization procedure, that was discussed in 
the bulk of our paper.

We have identified two different fixed points for our theory:  one
which we call the normal fixed point, and the other the Andreev fixed
point.  The normal fixed point is associated to the normal boundary
conditions (\ref{normalbcs}), and similarly for the Andreev fixed
point (\ref{Andreevbcs}).  At each of these fixed points,  it was shown that 
it is
possible to do the bosonization exactly in finite size.  
All correlation functions are then readily
computable, and we listed all correlators for all cases in Appendix A.  

We have presented explicit  computations for
physical quantities like the Josephson current, and the pairing order
parameter.  We summarize here these results, and perform a
stability analysis for the fixed points we have identified.

First, we define $I_{\parallel}$ and $I_{\perp}$ as the contributions to the 
Josephson
current from channel diagonal and channel off-diagonal correlators  
respectively.
Then we define the  set of operators  whose 
expectations values were already computed, and whose
scaling needs to be assessed.  
The channel-diagonal and channel
non-diagonal pairing operators are given by 
\begin{eqnarray}
P_{\parallel +} &=& \Psi_{L \downarrow}^i (x) \Psi_{L \uparrow}^i
(x), \hspace{2cm} P_{\parallel -} = \Psi_{L \downarrow}^i (x) \Psi_{R
\uparrow}^i (x), \nonumber \\ 
P_{\perp +} &=& \Psi_{L \downarrow}^i (x)
\Psi_{L \uparrow}^j (x), \hspace{2cm} P_{\perp -} = \Psi_{L \downarrow}^i (x)
\Psi_{R \uparrow}^j (x), 
\hspace{0.1cm} (i \neq j)
\label{pairings}
\end{eqnarray} 
both of which occur in the physical pairing operator involving the
full fermion.  For completeness , we  also define the
back scattering bilinears
\begin{eqnarray}
V_{\parallel +} &=& \Psi_{L \sigma}^{i \dagger} (x) \Psi_{L \sigma}^i
(x), \hspace{2cm} V_{\parallel -} = \Psi_{L \sigma}^{i \dagger} (x) \Psi_{R
\sigma}^i (x), \nonumber \\ 
V_{\perp +} &=& \Psi_{L \sigma}^{i \dagger} (x)
\Psi_{L \sigma}^j (x), \hspace{2cm} V_{\perp -} = \Psi_{L \sigma}^{i
\dagger} (x) \Psi_{R \sigma}^j (x), 
\hspace{0.1cm} (i \neq j).
\label{backs}
\end{eqnarray} 

In Tables \ref{tab:LLcoeff} and  \ref{tab:CNcoeff} we summarize the results 
obtained for the Josepshon current and pairing operator in the limits
of zero temperature and infinite size. For the  
diagonal and off-diagonal components of the 
Josephson current we list the exponent that determine their dependence
on the inverse of the system size, $\frac 1 R$. 
We also include the decay exponent of  the pairing operators as a
function of distance away from the boundary.
We list the results obtained considering the first 
non-vanishing contribution of the expectation value of any  
given operator, using the appropriate perturbation if necessary.

\begin{table}[H]
\caption{\label{tab:LLcoeff}Scaling exponents (Luttinger liquids).}
\begin{ruledtabular}
\begin{tabular}{lll}
  Operator & Normal Boundary Conditions & Andreev Boundary Conditions \\
\hline
  $I_{\parallel}$ &$ 2 e^{-2\xi_c^i} - 1$ & 1\\
  $I_{\perp}$ & $\sum_{a=c,s} \frac{1}{2}[e^{-2\xi_a^i} + e^{-2\xi_a^j}] -1$
& $\frac{1}{2}[ e^{2\xi_c^i} + e^{-2\xi_s^i} + (i \rightarrow j)] -1$ \\
$P_{\parallel,+}$ & $\frac{1}{2} [e^{2\xi_c^i} + 3 e^{-2\xi_c^i}] -1$  
& $\frac{1}{2}[3 e^{2\xi_c^i} + e^{-2\xi_c^i}] -1$ \\
$P_{\parallel,-}$ & $\frac{1}{2}[3 e^{-2\xi_c^i} +
e^{2\xi_s^i}] -1 $ &$\frac{1}{2}[e^{- 2\xi_c^i} + e^{ 2\xi^i_s}]$\\ 
$P_{\perp}$ & $\sum_{a} \frac{1}{8}[e^{2\xi_a^i} + 3 e^{-2\xi_a^i} + (i
\rightarrow j)] -1$
& $\frac{1}{8}[3 e^{2\xi_c^i} + e^{-2\xi_c^i} + e^{2\xi_s^i} + 3
e^{-2\xi_s^i} + (i \rightarrow j)] -1$\\ 
\end{tabular}
\end{ruledtabular}
\end{table}

\begin{table}[H]
\caption{\label{tab:CNcoeff}Scaling exponents (nanotube).}
\begin{ruledtabular}
\begin{tabular}{lll}
  Operator & Normal Boundary Conditions & Andreev Boundary Conditions \\
\hline
  $I_{\parallel}$ &$ e^{-2\xi_c}$ & 1\\
  $I_{\perp}$ & $e^{-2\xi_c}$ & $e^{2\xi_c}$ \\
$P_{\parallel,+}$ & $\frac{1}{4}[e^{2\xi_c} + 3 e^{-2\xi_c}]$ &
$\frac{1}{4}[3 e^{2\xi_c} + e^{-2\xi_c}]$ \\ 
$P_{\parallel,-}$ & $\frac{1}{4}[1 + 3e^{-2\xi_c}]$ & 
$\frac{1}{4}[e^{- 2\xi_c} + 3 ]$\\ 
$P_{\perp,+}$ & $\frac{1}{4}[e^{2\xi_c} + 3 e^{-2\xi_c}]$ &
$\frac{1}{4}[3 e^{2\xi_c} + e^{-2\xi_c} ]$ \\ 
$P_{\perp,-}$ & $\frac{1}{4}[1 + 3e^{-2\xi_c}]$ & 
$ \frac{1}{4}[e^{-2\xi_c} + 3] $\
\end{tabular}
\end{ruledtabular}
\end{table}

As it is clear from the previous section, in the case of Andreev boundary 
conditions,  the results listed for $I_{\parallel}$ and $P_{\parallel,-}$
 are exact, $I_{\perp}$ was obtained to second order in 
perturbation theory, and  $P_{\parallel,+}$ as well as $P_{\perp,\pm}$ 
were obtained to first order in perturbation theory. In all these cases the 
perturbative Hamiltonian corresponds to Andreev backscattering only.
For carbon nanotubes and Andreev boundary conditions, we included in the bulk 
of the paper only the off-diagonal terms of the boundary perturbation. 
In this case $<P_{\parallel,+}>=0$. However the result listed in table 
\ref{tab:CNcoeff} shows that this operator acquires a finite expectation value if 
the boundary perturbation includes diagonal terms. Finally, it is simple to 
see that the second order contribution to $P_{\perp,-}$ vanishes due the 
particular form of the boundary perturbation.  Therefore, althought we
include its naive exponent here, it should be remembered that in our
setup, this vanishes.  The scaling would then be determined by the
first descendant of this operator.

These tables are simply a convenient way of viewing the various dependencies
on the interaction parameters.  The full functional expression for the
operators' expectation values, however, is much more complicated,
containing nontrivial functions of the temperature, size of the
system, velocities and interaction parameters.  The results are too
bulky to summarize here more precisely than the table above, and we
refer the reader to the previous sections for the exact formulae.

What we have to do next is check the stability of our fixed points by
computing the effective dimension of the sets of
operators given by eqs. (\ref{pairings}) and (\ref{backs}), 
when bulk interactions are introduced.   We provide 
 four tables giving the scaling dimensions of the pairing and
backscattering 
operators around both the normal and Andreev fixed points for the
(multichannel) Luttinger liquids and nanotube.  
These exponents are all
simplified data coming from our exact correlators, and they are valid
for zero temperature in a large system.  
In the Luttinger liquid case, the
interaction parameters carry a channel index, whereas we have considered
here only one nanotube, with one charge interaction parameter.  

Three exponents are
given, corresponding to the various regimes of the correlators that
can be encountered.  $\Delta$ is the usual bulk scaling exponent,
valid away from the boundaries.  $\Delta_{\perp}$ is the scaling
exponent close to one boundary, with argument separation running
perpendicular to the boundary.  $\Delta_{\parallel}$ is the scaling
exponent for correlators where operators both live close to the
boundary, but with a finite time difference (the reader unfamiliar
with these various exponents is invited to see {\it e.g.}
\cite{MattsonPRB56} for a clear explanation).

To examine criteria of relevance in the RG sense in our theory, the
exponents to look at are in the $\Delta_{\parallel}$ columns.  That
is the case in our framework, since
our boundary perturbations include only fields sitting on the
boundaries, integrated over an infinite time interval;  in principle,
we should perform a 
complete RG analysis, including even terms which represent the
appearance of superconducting order within the one-dimensional
channel, away from the boundary.  We don't to this here, as these
contributions turn out to be much smaller.  It is readily seen that, for
normal boundary conditions, the pairing perturbations are irrelevant
($\Delta_{\parallel} < 1$) for repulsive interactions for both the
multichannel Luttinger liquid and the carbon nanotubes.  On the other
hand, the diagonal backscattering term is marginal (the cross-channel
backscattering is either irrelevant (Luttinger liquids) or marginal
(carbon nanotubes)).  Putting attractive interactions in the bulk,
however, destabilizes the normal fixed point:  in that case, the
operator that drives the system away is the channel-diagonal pairing
operator.  One therefore expects, on general grounds, in the case
of attractive bulk interactions (and supposing that our parameters are
initially tuned to put our system near the normal fixed point), to
witness a flow of our theory towards Andreev boundary conditions,
corresponding to an optimization of channel-diagonal Andreev
reflection at the boundary.  

At the Andreev fixed point, the operator scaling dimensions are
different, and are listed in the last two tables.  We stress this fact
again:  the particulars of the 
boundary conditions influence in a nontrivial way the scaling of the
system near the boundaries, and near the Andreev fixed point, the
dimensions of the operators differ from those to be found near the
normal fixed point.  The situation is now reversed as compared to the
scaling around the normal fixed point:  most operators are now
irrelevant for attractive interactions, with a few of them being
marginal.  If we were to turn the interactions in the bulk to the
repulsive regime, we would see the system flowing back to the normal
fixed point.  We therefore recover a similar situation to the one in
\cite{AffleckPRB62}, in which the normal fixed point was the fixed
point associated to repulsive bulk interactions, and the Andreev fixed
point was associated to attractive bulk interactions.  Our analysis
here is rather short and simplistic:  we could ask {\it e.g.} what
other possibilities there might be for flows in the system, in
particular by playing around with the interactions in the spin
channels.  A treatment similar to that in \cite{WongNPB417} would then
be obtained.  We leave this for future considerations.

The perturbative setup that we have presented here lends itself
ideally to further extensions and refinements.  First of all, the
formulas we have presented for the noninteracting case would be
appropriate to investigate the effects of voltages simultaneously
present with various backscattering potentials, or could easily be
modified for the addition of ferromagnetic couplings.  Other
possibilities more directly related to what we have done in the
interacting case would be, for example, to
take into account perturbatively the effects of quasiparticle
penetration within the original superconductors.  We have neglected
those according to the logic of \cite{AffleckPRB62}, and this should
be correct at extremely low temperatures.  However, finite
temperatures make quasiparticle penetration possible if the energy
scale starts approaching the (original, bulk) superconducting gap.
This could again be done within our framework, by replacing the
boundary action with some other action containing retarded effects.   
It is also important to realize that the class of problems that we have
dealt with here forms a good basis from which to study other geometries.
In view of the fact that experimental work is addressing these issues
on a frequent basis, we think that there exist many opportunities for
simple adaptations of our method.  One could 
consider things like nanotube crossings of various kinds, tunnel
junctions, etc. We leave such issues for later work.

\begin{table}[H]
\caption{\label{tab:LLcoeff1}Operator scaling dimensions
 (Luttinger liquids with normal boundary conditions).}
\begin{ruledtabular}
\begin{tabular}{llll}
  Operator &$\Delta$ & $\Delta_{\perp}$ & $\Delta_{\parallel}$ \\
\hline
$P_{\parallel,+}$ &$\cosh(2 \xi_c^i)$ &$\frac{1}{4}[e^{2\xi_c^i} +
3 e^{-2\xi_c^i}] $ & $e^{-2 \xi_c^i}$\\
$P_{\parallel,-}$ & $\frac{1}{2}[e^{-2\xi_c^i} + e^{2\xi_s^i}]$ 
&$\frac{1}{4}[3 e^{-2\xi_c^i} + e^{ 2\xi_s^i}]$ &$e^{-2 \xi_c^i}$\\ 
$P_{\perp , \pm}$ & $\frac{1}{4}[\cosh(2\xi_c^i) + \cosh(2\xi_s^i) + (i
\rightarrow j)]$
& $\frac{1}{16}[e^{2\xi_c^i} + 3e^{-2\xi_c^i} +e^{2\xi_s^i} + 3e^{-2\xi_s^i}+ (i
\rightarrow j)]$ & $\frac{1}{4}[e^{-2\xi_c^i} + e^{-2\xi_s^i} + (i
\rightarrow j)]$\\ 
$V_{\parallel,+}$ &$1$ &$1$ & $1$\\
$V_{\parallel,-}$ &$\frac{1}{2}[e^{2\xi_c^i} + e^{2\xi_s^i}]$ 
&$\frac{1}{4}[e^{2\xi_c^i} + e^{2\xi_s^i}+2]$ &$1$\\
$V_{\perp,\pm}$ &$\frac{1}{4}[\cosh(2\xi_c^i) + \cosh(2\xi_s^i) + (i
\rightarrow j)]$ 
&$  \frac{1}{16}[e^{2\xi_c^i} + 3e^{-2\xi_c^i} +e^{2\xi_s^i} + 3e^{-2\xi_s^i}+ (i
\rightarrow j)] $ &$\frac{1}{4}[e^{-2\xi_c^i} + e^{-2\xi_s^i} + (i
\rightarrow j)]$\\
\end{tabular}
\end{ruledtabular}
\end{table}

\begin{table}[H]
\caption{\label{tab:LLcoeff2}Operator scaling dimensions
(Luttinger liquids with Andreev boundary conditions).}
\begin{ruledtabular}
\begin{tabular}{llll}
  Operator &$\Delta$ & $\Delta_{\perp}$ & $\Delta_{\parallel}$ \\
\hline
$P_{\parallel,+}$ &$\cosh(2 \xi_c^i)$ &$\frac{1}{4}[3 e^{2\xi_c^i} +
e^{-2\xi_c^i}] $ & $e^{2 \xi_c^i}$\\
$P_{\parallel,-}$ & $\frac{1}{2}[e^{-2\xi_c^i} + e^{2\xi_s^i}]$ 
&$\frac{1}{4}[e^{-2\xi_c^i} + e^{2\xi_s^i}+2]$ &$1$\\ 
$P_{\perp , \pm}$ & $\frac{1}{4}[\cosh(2\xi_c^i) + \cosh(2\xi_s^i) + (i
\rightarrow j)]$
& $\frac{1}{16}[3 e^{2\xi_c^i} + e^{-2\xi_c^i} + 
e^{2\xi_s^i} + 3e^{-2\xi_s^i} +
(i \rightarrow j)]$ & $\frac{1}{4}[e^{2\xi_c^i} + e^{-2\xi_s^i} + (i
\rightarrow j)]$\\
$V_{\parallel,+}$ &$1$ &$1$ & $1$\\
$V_{\parallel,-}$ &$\frac{1}{2}[e^{2\xi_c^i} + e^{2\xi_s^i}]$ 
&$\frac{1}{4}[3 e^{2\xi_c^i} + e^{2\xi_s^i}]$ &$e^{2\xi_c^i}$\\
$V_{\perp,\pm}$ &$\frac{1}{4}[\cosh(2\xi_c^i) + \cosh(2\xi_s^i) + (i
\rightarrow j)]$ 
&$  \frac{1}{16}[3 e^{2\xi_c^i} + e^{-2\xi_c^i} + e^{2\xi_s^i} + 3 e^{-2\xi_s^i}+ (i
\rightarrow j)] $ &$\frac{1}{4}[e^{2\xi_c^i} + e^{-2\xi_s^i} + (i
\rightarrow j)]$\\ 
\end{tabular}
\end{ruledtabular}
\end{table}

\begin{table}[H]
\caption{\label{tab:CNcoeff1}Operator scaling dimensions 
(nanotube with normal boundary conditions).}
\begin{ruledtabular}
\begin{tabular}{llll}
  Operator &$\Delta$ & $\Delta_{\perp}$ & $\Delta_{\parallel}$ \\
\hline
$P_{\parallel,+}$ &$\frac{1}{2}[\cosh(2\xi_c) +1] $ &
$\frac{1}{8}[e^{2\xi_c} + 3e^{-2\xi_c} +4]$ & 
$\frac{1}{2}[e^{-2 \xi_c}+1]$\\
$P_{\parallel,-}$ & $\frac{1}{4}[e^{-2\xi_c} + 3]$ 
&$\frac{1}{8}[3 e^{-2\xi_c} + 5]$ &$\frac{1}{2}[e^{-2 \xi_c}+1]$\\ 
$P_{\perp , +}$ & $\frac{1}{2}[\cosh(2\xi_c) + 1]$
& $\frac{1}{8}[e^{2\xi_c} + 3e^{-2\xi_c} + 4]$ & 
$\frac{1}{2}[e^{-2\xi_c} + 1]$\\
$P_{\perp , -}$ & $\frac{1}{4}[e^{-2\xi_c} + 3]$
& $\frac{1}{8}[3 e^{-2\xi_c} + 5]$ & 
$\frac{1}{2}[e^{-2\xi_c} + 1]$\\ 
$V_{\parallel,\perp,+}$ &$1$ &$1$ & $1$\\
$V_{\parallel,\perp,-}$ &$\frac{1}{4}[e^{2\xi_c} + 3]$ 
&$\frac{1}{8}[ e^{2\xi_c} + 7]$ &$1$\\
\end{tabular}
\end{ruledtabular}
\end{table}

\begin{table}[H]
\caption{\label{tab:CNcoeff2}Operator scaling dimensions 
(nanotubes with Andreev boundary conditions).}
\begin{ruledtabular}
\begin{tabular}{llll}
  Operator &$\Delta$ & $\Delta_{\perp}$ & $\Delta_{\parallel}$ \\
\hline
$P_{\parallel,+}$ &$\frac{1}{2}[\cosh(2\xi_c) +1] $ &
$\frac{1}{8}[3e^{2\xi_c} + e^{-2\xi_c}+4]$ & 
$\frac{1}{2}[e^{2 \xi_c}+1]$\\

$P_{\parallel,-}$ & $\frac{1}{4}[e^{-2\xi_c} + 3]$ 
&$\frac{1}{8}[e^{-2\xi_c} + 7]$ &$1$\\ 

$P_{\perp , +}$ & $\frac{1}{2}[\cosh(2\xi_c) + 1]$
& $\frac{1}{8}[3e^{2\xi_c} + e^{-2\xi_c} + 4]$ & 
$\frac{1}{2}[e^{2\xi_c} + 1]$\\

$P_{\perp , -}$ & $\frac{1}{4}[e^{-2\xi_c} + 3]$
& $\frac{1}{8}[ e^{-2\xi_c} + 7]$ & 
$ 1 $\\
$V_{\parallel,\perp,+}$ &$1$ &$1$ & $1$\\
$V_{\parallel,\perp,-}$ &$\frac{1}{4}[e^{2\xi_c} + 3]$ 
&$\frac{1}{8}[3 e^{2\xi_c} + 5]$ &${\frac 1 2 }[e^{2\xi_c}+1]$\\
\end{tabular}
\end{ruledtabular}
\end{table}

\appendix

\section{Correlation functions}
\label{apA}

The usefulness of bosonization really manifests itself in the
computation of correlation functions.  In this appendix, we compute  
multipoint correlation functions, in finite size and
at finite temperature, for all the cases of interest in this work.

The first requirement when dealing with a finite size is to specify
the boundary conditions used in the computation. We consider here normal 
and Andreev boundary conditions introduced
in the bulk of the paper.  We present computations both for the
Luttinger liquid, and for the carbon nanotubes.  Although two-point 
correlators for such systems are already known in finite size and at
finite temperatures, and multipoint correlators in the bulk are also
in the literature, we wish here to present formulas for any multipoint
correlator.

\subsection{Luttinger liquid, normal boundary conditions}

Let us consider the problem of computing correlation functions in
finite size, at finite temperature, for normal boundary conditions.  
The general correlator that we want to compute can be taken to be 
\begin{eqnarray}
G^{(2n)}_{\{ \sigma, \tau \} } (\{ x, t; y, u\}) 
\equiv \langle \prod_{i=1}^n \Psi_{L
\sigma_i}^{\dagger} (x_i, t_i) \prod_{j=1}^n \Psi_{L \tau_j} (y_j, u_j)
\rangle_N
\end{eqnarray}
since all others can be obtained by simple manipulations (here, we
mean in particular correlators with a different ordering of the $\Psi,
\Psi^{\dagger}$ operators, which are easy to recover from the formulas
we present;  see later
for a better explanation).  Correlators involving right movers can be
recovered by
rewriting the right movers as left movers 
using the appropriate (here, normal) boundary conditions.  
Here, the angular brackets denote an average
using the Hamiltonian (\ref{NBCsHamiltonian}), and the subindex $N$
denotes normal boundary conditions.  

For simplicity, we have suppressed the channel index. We remark that
when dealing with multichannel LL, all the correlators needed to
compute various physical quantities can be reduced to the above one
for a given channel , or products of them evaluated in different
channels. 

The first step in the computation is to substitute the bosonization formulas
(\ref{LLBosonization}) for the fermions, and to make use of the normal
boundary conditions, mode expansions and Hamiltonian (equations (\ref{NBCs})
to (\ref{NBCsHamiltonian})) to reexpress the correlator as a product of zero-mode
and dynamical contributions, averaged over a free theory.  We write this as
\begin{eqnarray}
G^{(2n)}_{\{ \sigma, \tau \} } (\{ x, t; y, u\}) = \delta_{\sum
\sigma, \sum \tau}
J_{\{ \sigma, \tau \} } (\{ x, t; y, u\}) K_{\{ \sigma, \tau \} } (\{ x, t; y, u\}).
\end{eqnarray}

Let us first consider the zero-mode part.  Using the Campbell-Baker-Hausdorff
formula to combine exponentials, we can reexpress this contribution as
\begin{eqnarray}
J_{\{ \sigma, \tau \} } (\{ x, t; y, u\}) = 
P_c (\{ x, t; y, u\}) P_{s \{ \sigma, \tau \} } (\{ x, t; y, u\})
Q_{\{ \sigma, \tau \} } (\{ x, t; y, u\})
\end{eqnarray}
with phase functions 
\begin{eqnarray}
\ln P_c = \frac{-i \pi}{4R} \left( n \sum_{i=1}^n [v_c e^{-2 \xi_c} (t_i - u_i)
+ x_i - y_i] - \sum_{i < j = 1}^n [v_c e^{-2 \xi_c} (t_i - t_j + u_i - u_j) 
+ x_i - x_j + y_i - y_j] \right), \nonumber \\
\ln P_{s \{ \sigma, \tau \} } (\{ x, t; y, u\}) = 
 \frac{-i \pi}{4R} \left( [\sum_i \sigma_i] \sum_{i=1}^n [\sigma_i (v_s e^{-2 \xi_s} t_i 
+ x_i) - \tau_i (v_s e^{-2\xi_s} u_i + y_i)] - \right. \nonumber \\
\left. - \sum_{i<j =1}^n [\sigma_i \sigma_j
(v_s e^{-2\xi_s} (t_i - t_j) + x_i - x_j) + \tau_i \tau_j (v_s e^{-2\xi_s} (u_i - u_j)
+ y_i - y_j) ] \right)
\end{eqnarray}
and zero-mode expectation value
\begin{eqnarray}
Q_{\{ \sigma, \tau \} } (\{ x, t; y, u\}) = \langle e^{i \sqrt{2\pi} \frac{\hat{\Pi}_c}{2R} 
\sum_{i=1}^n [v_c e^{-\xi_c} (t_i - u_i) + e^{\xi_c} (x_i - y_i)]}
e^{i \sqrt{2\pi} \frac{\hat{\Pi}_s}{2R} \sum_{i=1}^n [ v_s e^{-\xi_s} (\sigma_i t_i 
- \tau_i u_i) + e^{\xi_s} (\sigma_i x_i - \tau_i y_i)]} \rangle.
\end{eqnarray} 
This average is readily computed in terms of $\theta$-functions (see next appendix for
definitions and useful formulas).  The final result is given below in equation
(\ref{NBCzeromodepart}).  Note the crucial fact that the spin and charge sectors
remain entangled in this contribution.

The dynamical part, on the other hand, readily factorizes in charge and spin sectors:
\begin{eqnarray}
K_{\{ \sigma, \tau \} } (\{ x, t; y, u\}) = K_c (\{ x, t; y, u\}) 
K_{s \{ \sigma, \tau \} } (\{ x, t; y, u\}).
\end{eqnarray}
For clarity, we will here present details of the computation of the spin part
only (the charge part doesn't depend on the $\sigma, \tau$ subindices and is
simple to recover from the formulas below).  As an expectation value, this is
\begin{eqnarray}
K_{s \{ \sigma, \tau \} } (\{ x, t; y, u\}) = 
\langle \prod_{i=1}^n e^{i \sqrt{2\pi} \sigma_i [ \cosh \xi_s \: \zeta_{s L} (x_i, t_i)
- \sinh \xi_s \: \zeta_{s L} (-x_i, t_i)]} 
\prod_{i=1}^n e^{-i \sqrt{2\pi} \tau_i [ \cosh \xi_s \: \zeta_{s L} (y_i, u_i)
- \sinh \xi_s \: \zeta_{s L} (-y_i, u_i)]} \rangle
\end{eqnarray}
Since the average is over a free bosonic field, we can make use of the identity
\begin{eqnarray}
\langle \prod_i e^{A_i} \rangle = e^{\langle \left[ \sum_{i < j} A_i A_j + \frac{1}{2}
\sum_i A_i^2 \right] \rangle}
\end{eqnarray}
to rewrite the above expectation value as
\begin{eqnarray}
\ln K_{s \{ \sigma, \tau \} } (\{ x, t; y, u\}) = \hspace{12cm} \nonumber \\
=
2\pi \cosh^2 \xi_s [ \sum_{i,j} \sigma_i \tau_j B_s (x_i, t_i; y_j, u_j)
- \sum_{i<j} \sigma_i \sigma_j B_s (x_i, t_i; x_j, t_j) - \sum_{i<j} \tau_i \tau_j
B_s (y_i, u_i; y_j, u_j)] + \nonumber \\
+ 2\pi \sinh^2 \xi_s  [ \sum_{i,j} \sigma_i \tau_j B_s (-x_i, t_i; -y_j, u_j)
- \sum_{i<j} \sigma_i \sigma_j B_s (-x_i, t_i; -x_j, t_j) - \sum_{i<j} \tau_i \tau_j
B_s (-y_i, u_i; -y_j, u_j)] - \nonumber \\
- \pi \sinh 2\xi_s  [ \sum_{i,j} \sigma_i \tau_j (B_s (x_i, t_i; -y_j, u_j)
+ B_s (-x_i, t_i; y_j, u_j))
- \nonumber \\ 
- \sum_{i<j} \sigma_i \sigma_j (B_s (x_i, t_i; -x_j, t_j) + B_s (-x_i, t_i; x_j, t_j))
- \sum_{i<j} \tau_i \tau_j (B_s (y_i, u_i; -y_j, u_j) +B_s (-y_i, u_i;
y_j, u_j))
- \nonumber \\
- \frac{1}{2} \sum_i (B_s(x_i, t_i; -x_i, t_i) + B_s(-x_i, t_i; x_i, t_i) +
B_s (y_i, u_i; -y_i, u_i) + B_s (-y_i, u_i; y_i, u_i)) ]
\end{eqnarray}
where 
\begin{eqnarray}
B_a (x,t; y,u) = \langle \zeta_{a L} (x,t) \zeta_{a L} (y,u) \rangle
- \frac{1}{2} \langle \zeta_{a L} (x,t) \zeta_{a,L} (x,t) \rangle
- \frac{1}{2} \langle \zeta_{a L} (y,u) \zeta_{a,L} (y,u) \rangle.
\end{eqnarray}
The computation of this free correlator can also be done in terms of
$\theta$-functions, with the result
\begin{eqnarray}
B_a (x,t; y,u) = \frac{i}{4} \frac{v_a (t - u) + x - y}{2R} 
- \frac{1}{4\pi} \ln F_a (v_a (t-u) + x-y)
\end{eqnarray}
where
\begin{eqnarray}
F_a (z) = \frac{\vartheta_1 (\frac{\pi z}{2R} | \omega_a)}
{\vartheta_1 (-i \frac{\pi \alpha}{2R} | \omega_a)} 
\end{eqnarray}
in which $\alpha$ is an implicit cutoff, i.e. $z \rightarrow z - i 
\alpha$.  The periods of the theta functions are
\begin{eqnarray}
\omega_a = i \frac{v_a \beta}{2R}.
\end{eqnarray}

This whole collection of formulas can now be assembled to yield our
original correlator.  In doing this, phases simplify somewhat;
in the end, we obtain the general formula
\begin{eqnarray}
G^{(2n)}_{\{\sigma, \tau \} } (\{ x, t; y, u\}) = {\cal N}_n^{-1}
\delta_{\sum_i \sigma_i, 
\sum_i \tau_i} Q_{ \{\sigma, \tau \} } (\{ x, t; y, u\}) \times
\nonumber \\
\times
\prod_{a = c,s} \left[ P_{a + -} \right]^{-\frac{\cosh^2
\xi_a}{2}} \left[ P_{a - +} \right]^{- \frac{\sinh^2 \xi_a}{2}}
\left[ P_{a ++} P_{a --} \right]^{\frac{\sinh 2\xi_a}{4}}
\prod_{i=1}^n |F_a (2x_i) F_a (2y_i)|^{-\frac{\sinh 2\xi_a}{4}}
\end{eqnarray}
where for $\eta_i = \pm$ we have defined the functions
\begin{eqnarray}
P_{c \eta_1 \eta_2}  = \frac{\prod_{i,j = 1}^n F_c (v_c (t_i - u_j) + \eta_1 x_i
+ \eta_2 y_j)}{\prod_{i<j=1}^n F_c (v_c(t_i - t_j)
+ \eta_1 x_i + \eta_2 x_j) F_c (v_c (u_i - u_j) + \eta_1 y_i + \eta_2
y_j)}, \nonumber \\
P_{s \eta_1 \eta_2} = \frac{\prod_{i,j=1}^n F_s^{\sigma_i \tau_j} (v_s (t_i -
u_j) + \eta_1 x_i + \eta_2 y_j)}{\prod_{i<j=1}^n F_s^{\sigma_i
\sigma_j} (v_s(t_i - t_j) 
+ \eta_1 x_i + \eta_2 x_j) F_s^{\tau_i \tau_j} (v_s (u_i - u_j) +
\eta_1 y_i + \eta_2 y_j)},
\end{eqnarray}
and where the zero mode part is
\begin{eqnarray}
Q_{ \{ \sigma, \tau \} } (\{x, t; y,u \}) = \frac{\vartheta_2 (d_c^t +
\tau_c k_F R | \tau_c) 
\vartheta_3 (d_s^t | \tau_s) + \vartheta_3 (d_c^t + \tau_c k_F R | \tau_c)
\vartheta_2 (d_s^t | \tau_s)}{\vartheta_2 (\tau_c k_F R | \tau_c)
\vartheta_3 (0 | \tau_s) + \vartheta_3 (\tau_c k_F R | \tau_c)
\vartheta_2 (0 | \tau_s)} e^{i 2 d_c^t k_F R/\pi}
\label{NBCzeromodepart}
\end{eqnarray}
in which we have defined $\tau_a = i \frac{\beta v_a e^{-2\xi_a}}{R}$ and
\begin{eqnarray}
d_c^t = -\frac{\pi}{2R} \sum_{i = 1}^n 
\left[ e^{-2 \xi_c} v_c (t_i - u_i) + x_i - y_i \right], \nonumber \\
d_s^t = -\frac{\pi}{2R} \sum_{i=1}^n \left[ e^{-2 \xi_s} v_s
(\sigma_i t_i - \tau_i u_i) + \sigma_i x_i - \tau_i y_i \right].
\end{eqnarray}
One can convince oneself that the correlators obey the correct
(anti-)periodicity by shifting imaginary time like in the Matsubara
formalism, and that moreover they obey the spatial periodicity by
shifting distance arguments by twice the system size.  This will of
course be true of all correlators we present, in all differing
boundary conditions cases.

Other correlators, for example of the type where the order of the
$\Psi, \Psi^{\dagger}$ differs, can be easily obtained from the above
formulas:  the only differences with the formulas that we have
presented are some simple signs, in front of the coordinates and
exponents.  We have left these out here to keep the formulas more
compact, and since these modifications are trivial to work out.

\subsection{Luttinger liquid, Andreev boundary conditions}

The general correlator is here again defined as
\begin{eqnarray}
\tilde{G}^{(2n)}_{\{ \sigma, \tau \} } (\{ x, t; y, u\}) 
\equiv \langle \prod_{i=1}^n \Psi_{L
\sigma_i}^{\dagger} (x_i, t_i) \prod_{j=1}^n \Psi_{L \tau_j} (y_j, u_j)
\rangle_A
\end{eqnarray}
where we now deal with Andreev boundary conditions, specified by the
subindex $A$.  The computation is rather similar to the one above;  in
fact, one can convince oneself rather easily that the general formula
becomes a simple variation of the one for normal boundary conditions, i.e.
\begin{eqnarray}
\tilde{G}^{(2n)}_{\{\sigma, \tau \} } (\{ x, t; y, u\}) = {\cal N}_n^{-1}
\delta_{\sum_i \sigma_i, 
\sum_i \tau_i} \tilde{Q}_{ \{\sigma, \tau \} } (\{ x, t; y, u\}) \times
\nonumber \\
\times
\left[ \prod_{a = c,s} \left[ P_{a + -} \right]^{-\frac{\cosh^2
\xi_a}{2}} \left[ P_{a - +} \right]^{- \frac{\sinh^2 \xi_a}{2}} \right]
\left[ P_{c ++} P_{c --} \right]^{-\frac{\sinh 2\xi_c}{4}}
\left[ P_{s ++} P_{s --} \right]^{\frac{\sinh 2\xi_s}{4}}
\prod_{i=1}^n 
\frac{
|F_c (2x_i) F_c (2y_i)|^{\frac{\sinh 2\xi_c}{4}}}{|F_s (2x_i) F_s
(2y_i)|^{\frac{\sinh 2\xi_s}{4}}} 
\end{eqnarray}
where the $P$ functions are as for normal boundary conditions, and
where the zero mode part is 
\begin{eqnarray}
\tilde{Q}_{ \{ \sigma, \tau \} } (\{x, t; y,u \}) = \frac{\vartheta_2
(\tilde{d}_c^t + \tilde{\tau}_c \chi/2 | \tilde{\tau}_c) 
\vartheta_2 (d_s^t | \tau_s) + \vartheta_3 (\tilde{d}_c^t + \tilde{\tau}_c
\chi/2 | \tilde{\tau}_c)
\vartheta_3 (d_s^t | \tau_s)}{\vartheta_2 (\tilde{\tau}_c \chi/2 |
\tilde{\tau}_c) 
\vartheta_2 (0 | \tau_s) + \vartheta_3 (\tilde{\tau}_c \chi/2 |
\tilde{\tau}_c) 
\vartheta_3 (0 | \tau_s)} e^{i \tilde{d}_c^t \chi/\pi}
\label{ABCzeromodepart}
\end{eqnarray}
in which we have defined $\tilde{\tau}_c = i \frac{\beta v_c
e^{2\xi_c}}{R}$ and 
\begin{eqnarray}
\tilde{d}_c^t = -\frac{\pi}{2R} \sum_{i = 1}^n 
\left[ e^{2 \xi_c} v_c (t_i - u_i) + x_i - y_i \right].
\end{eqnarray}  
The other parameters are defined as in the previous section.

\subsection{Carbon nanotube, normal boundary conditions}

The general correlator for carbon nanotubes in the presence of normal
boundary conditions can be written in terms of 
\begin{eqnarray}
G^{\{s,r \}}_{\{ \sigma, \tau \} } (\{ x, t; y, u\}) 
\equiv \langle \prod_{i=1}^n \Psi_{L
\sigma_i}^{s_i \dagger} (x_i, t_i) \prod_{j=1}^n \Psi_{L \tau_j}^{r_j}
(y_j, u_j) 
\rangle_N
\end{eqnarray}
where $s_i, r_j = 1,2$ refer to the nanotube channel index.  The
computation of this general correlator is very similar to the
computation presented above for a Luttinger liquid with normal
boundary conditions.  The final formula is
\begin{eqnarray}
G^{\{s,r \}}_{\{ \sigma, \tau \} } (\{ x, t; y, u\}) = {\cal N}_n^{-1}
\delta_{Klein} Q_{\{\sigma, \tau \}}^{\{ s,r \}} (\{ x,t;y,u\}) 
\times \nonumber \\ 
\times \left[ P_{c+-}^+ \right]^{-\frac{\cosh^2
\xi_c}{4}} \left[ P_{c-+}^+ \right]^{-\frac{\sinh^2 \xi_c}{4}} \left
[ P_{c++}^+ P_{c--}^+ \right]^{\frac{\sinh 2\xi_c}{8}} \left
[ P_{c+-}^- P_{s+-}^+ P_{s+-}^- \right]^{-1/4} \prod_{i=1}^n |F_c
(2x_i) F_c (2y_i)|^{-\frac{\sinh 2\xi_c}{8}}.
\end{eqnarray}
In the above, $\delta_{Klein}$ is a shorthand for the Kronecker deltas
coming from the Klein factors, meaning that the set of (pairs of) indices
associated to the $\Psi^{\dagger}$ operators has to be a permutation
of the set associated to the $\Psi$ operators.  For later convenience,
we also define $s(s_i)$ as $+1$ for the
$1$ channel, $-1$ for the $2$ channel (and similarly for $s(r_j)$).
We have defined the dynamical contributions
\begin{eqnarray}
P_{a \eta_1 \eta_2}^{\pm} = \frac{\prod_{i,j=1}^n [F_a (v_a^{\pm} (t_i - u_j) +
\eta_1 x_i +\eta_2 y_j)]^{\epsilon_a^{\pm}(i) \gamma_a^{\pm}(j)}}{\prod_{i<j=1}^n
[F_a (v_a^{\pm} (t_i - t_j) + \eta_1 x_i +\eta_2
x_j)]^{\epsilon_a^{\pm}(i) \epsilon_a^{\pm}(j)} [F_a (v_a^{\pm} (u_i - u_j)
+ \eta_1 y_i + \eta_2 y_j)]^{\gamma_a^{\pm}(i) \gamma_a^{\pm}(j)}}, 
\end{eqnarray}
with indices
\begin{eqnarray}
\epsilon_c^+ (i) = 1, \hspace{1cm} \epsilon_c^-(i) = s(s_i), 
\hspace{1cm} \epsilon_s^+ (i) = \sigma_i, \hspace{1cm} \epsilon_s^-
(i) = s(s_i) \sigma_i, \nonumber \\
\gamma_c^+ (j)= 1, \hspace{1cm} \gamma_c^- (j) = s(r_j), \hspace{1cm}
\gamma_s^+ (j)= \tau_j, \hspace{1cm} \gamma_s^- (j)= s(r_j) \tau_j.
\end{eqnarray}
and $v_a^+=v_a^-=v_a$, with $a=c,s$.
The zero mode part is
\begin{eqnarray}
Q_{\{\sigma, \tau \}}^{\{ s,r \}} (\{ x,t;y,u\}) =
\frac{\theta(\mathbf{u} + \frac{\tau_c}{2} (k_F R - \frac{\pi}{2}) |
\mathbf{\Omega} )}{\theta(\frac{\tau_c}{2} (k_F R - \frac{\pi}{2}) |
\mathbf{\Omega} )} e^{-i d_c^+ (1 - \frac{2 k_F R}{\pi})},
\end{eqnarray}
where $\mathbf{\Omega}$ is a four-dimensional $\theta$-function
(such multivariable functions are defined in the next appendix).  The
four-dimensional period matrix is ($\sigma^1$ is the Pauli matrix)
\begin{eqnarray}
\mathbf{\Omega} = i\frac{\beta}{4R} \left[ \mathbf{1} \otimes (\mathbf
1 + \mathbf{\sigma}^1) \frac{v_c}{2}(1 + e^{-2\xi_c}) -
\mathbf{\sigma}^1 \otimes (\mathbf{1} + \mathbf{\sigma}^1)
\frac{v_c}{2} (1 - e^{-2\xi_c}) + \mathbf{1} \otimes (\mathbf{1} -
\mathbf{\sigma}^1) v_s \right]
\label{Omega}
\end{eqnarray}
and $\tau_c = i\frac{\beta v_c e^{-2\xi_c}}{R}$.  The vector
$\mathbf{u}$ is
\begin{eqnarray}
\mathbf{u} = \frac{1}{4} \left( 
\begin{array}{c}
d_c^+ + d_c^- + d_s^+ + d_s^- \\
d_c^+ + d_c^- - d_s^+ - d_s^- \\
d_c^+ - d_c^- + d_s^+ - d_s^- \\
d_c^+ - d_c^- - d_s^+ + d_s^- 
\end{array} \right),
\end{eqnarray}
in which the distances are given by
\begin{eqnarray}
d_a^{\pm} = -\frac{\pi}{2R} \sum_{i = 1}^n 
\left[ e^{-2 \xi_a^{\pm}} v_a (\epsilon_a^{\pm} t_i - \gamma_a^{\pm}
u_i) + \epsilon_a^{\pm} x_i - \gamma_a^{\pm} y_i \right].
\end{eqnarray}  
Also, we put $\xi_c^- =\xi_s^\pm = 0$, $\xi_c^+=\xi_c$ 
and  $k_F^1 = k_F^2 \equiv k_F$.

\subsection{Carbon nanotube, Andreev boundary conditions}

For Andreev boundary conditions, we define 
\begin{eqnarray}
\tilde{G}^{\{s,r \}}_{\{ \sigma, \tau \} } (\{ x, t; y, u\}) 
\equiv \langle \prod_{i=1}^n \Psi_{L
\sigma_i}^{s_i \dagger} (x_i, t_i) \prod_{j=1}^n \Psi_{L \tau_j}^{r_j} (y_j, u_j)
\rangle_A
\end{eqnarray}
where $s_i, r_j = 1,2$ refer to the nanotube channel index.  
The computation is very similar to those above, and yields
\begin{eqnarray}
\tilde{G}^{\{s,r \}}_{\{ \sigma, \tau \} } (\{ x, t; y, u\}) = {\cal N}_n^{-1}
\delta_{Klein} \tilde{Q}_{\{\sigma, \tau \}}^{\{ s,r \}} (\{ x,t;y,u\}) 
\times \nonumber \\
 \left[ P_{c+-}^+ \right]^{-\frac{\cosh^2
\xi_c}{4}} \left[ P_{c-+}^+ \right]^{-\frac{\sinh^2 \xi_c}{4}} \left
[ P_{c++}^+ P_{c--}^+ \right]^{-\frac{\sinh 2\xi_c}{8}} \left
[ P_{c+-}^- P_{s+-}^+ P_{s+-}^- \right]^{-1/4} \prod_{i=1}^n |F_c
(2x_i) F_c (2y_i)|^{\frac{\sinh 2\xi_c}{8}}
\end{eqnarray}
The zero mode part becomes here
\begin{eqnarray}
\tilde{Q}_{\{\sigma, \tau \}}^{\{ s,r \}} (\{ x,t;y,u\}) =
\frac{\theta(\tilde{\mathbf{u}} + \frac{\tilde{\tau}_c \chi}{4} |
\tilde{\mathbf{\Omega}} )}{\theta(\frac{\tilde{\tau}_c \chi}{4} |
\tilde{\mathbf{\Omega}} )} e^{i \tilde{d}_c^+ \chi/\pi}
\end{eqnarray}
with period matrix
\begin{eqnarray}
\tilde{\mathbf{\Omega}} = i\frac{\beta}{4R} \left[ \mathbf{1} \otimes (\mathbf
1 + \mathbf{\sigma}^1) \frac{v_c}{2}(1 + e^{2\xi_c}) -
\mathbf{\sigma}^1 \otimes (\mathbf{1} + \mathbf{\sigma}^1)
\frac{v_c}{2} (1 - e^{2\xi_c}) + \mathbf{1} \otimes (\mathbf{1} -
\mathbf{\sigma}^1) v_s \right]
\end{eqnarray}
and $\tilde{\tau}_c = i\frac{\beta v_c e^{2\xi_c}}{R}$.  The vector
$\tilde{\mathbf{u}}$ is
\begin{eqnarray}
\tilde{\mathbf{u}} = \frac{1}{4} \left( 
\begin{array}{c}
\tilde{d}_c^+ + d_c^- + d_s^+ + d_s^- \\
\tilde{d}_c^+ + d_c^- - d_s^+ - d_s^- \\
\tilde{d}_c^+ - d_c^- + d_s^+ - d_s^- \\
\tilde{d}_c^+ - d_c^- - d_s^+ + d_s^- 
\end{array} \right).
\end{eqnarray}
The newly defined distances (only $\tilde{d}_c^+$ is actually needed,
as in this particular setup the others coincide with the previous ones)
are given by 
\begin{eqnarray}
\tilde{d}_a^{\pm} = -\frac{\pi}{2R} \sum_{i = 1}^n 
\left[ e^{2 \xi_a^{\pm}} v_a (\epsilon_a^{\pm} t_i - \gamma_a^{\pm}
u_i) + \epsilon_a^{\pm} x_i - \gamma_a^{\pm} y_i \right].
\end{eqnarray}  
We have also put $\chi_1 = \chi_2 \equiv \chi$.

\section{$\theta$-functions, and limits of correlation functions}

\subsection{Traditional $\theta$-functions}

Theta functions are defined with the following series:
\begin{eqnarray}
&\theta_1 (u | \tau) = -i \sum_{n \in \mathbb{Z}} (-1)^n 
e^{i \pi \tau (n+1/2)^2 + i (2 n+1) u}, \hspace{3cm} 
&\theta_2 (u | \tau) = \sum_{n \in \mathbb{Z}}  e^{i \pi \tau (n+ 1/2)^2 
+ i (2n+1) u}, \nonumber \\
&\theta_3 (u | \tau) = \sum_{n \in \mathbb{Z}} e^{i \pi \tau n^2 
+ i 2 n u}, \hspace{3cm} 
&\theta_4 (u | \tau) = \sum_{n \in \mathbb{Z}} (-1)^n e^{i \pi \tau n^2 
+ i 2n u}. 
\end{eqnarray}
The most important property of these functions is their quasi-double-
periodicity in the complex plane.  We refer the reader to the standard
literature for a list of all the properties obeyed by these remarkable
functions \cite{theta}.

In taking different limits like that of finite size and/or zero temperature,
we need to consider various limits of theta functions with respect to 
their arguments and periods.  We here collect a few useful formulas:
\begin{eqnarray}
&&\theta_1 (u |i \alpha) \rightarrow_{\alpha \rightarrow \infty} 2
e^{-\pi \alpha/4}  
\sin(u) + ... \nonumber \\
&&\theta_1 (u |i \alpha) \rightarrow_{\alpha \rightarrow 0} 2 \alpha^{-1/2}
e^{-\frac{\pi}{4\alpha} - \frac{u^2}{\pi\alpha}} \sinh \left(
\frac{u}{\alpha}\right)+ ... \nonumber \\
&&\theta_2 (u |i \alpha ) \rightarrow_{\alpha \rightarrow \infty} 
2 e^{-\pi \alpha/4} \cos(u) + ... \nonumber \\
&&\theta_2 (u |i \alpha ) \rightarrow_{\alpha \rightarrow 0} 
\alpha^{-1/2} e^{-\frac{u^2}{\pi \alpha}} \left[ 1 - 2 e^{-\pi/\alpha}
\cosh \left(\frac{2u}{\alpha}\right) \right] + ... \nonumber \\
&&\theta_3 (u |i \alpha ) \rightarrow_{\alpha \rightarrow \infty} 
1 + 2 e^{-\pi \alpha} \cos(2u) + ... \nonumber \\
&&\theta_3 (u |i \alpha ) \rightarrow_{\alpha \rightarrow 0} 
\alpha^{-1/2} e^{-\frac{u^2}{\pi \alpha}} \left[ 1 + 2 e^{-\pi/\alpha}
\cosh \left(\frac{2u}{\alpha}\right) \right] + ...\nonumber \\ 
&&\theta_4 (u |i \alpha ) \rightarrow_{\alpha \rightarrow \infty} 
1 - 2 e^{-\pi \alpha} \cos(2u) + ... \nonumber \\
&&\theta_4 (u |i \alpha ) \rightarrow_{\alpha \rightarrow 0} 
\alpha^{-1/2} e^{-\frac{u^2}{\pi \alpha}} \left[ 2 e^{-\pi/ 4\alpha}
\cosh \left(\frac{u}{\alpha}\right) \right] + ... 
\end{eqnarray}
where $...$ denote subdominant terms.

\subsection{Multivariable theta functions}

For the nanotube, the above $\theta$-functions are not sufficient.  It
is rather more convenient to use multivariable functions
\cite{theta}, 
defined as
\begin{eqnarray}
\theta (\mathbf{z} | \mathbf{\Omega}) = \sum_{\mathbf{n} \in
\mathbb{Z}^g} e^{i \pi \mathbf{n}^t \mathbf{\Omega} \mathbf{n} + 2i
\mathbf{n}^t \mathbf{z} }
\end{eqnarray}
where $g$ is an integer (the ``dimensionality'' of the function),
$\mathbf{\Omega}$ is a $g \times g$ matrix with positive imaginary parts, and
the sum is taken over all $g$-dimensional real vectors with integer
coefficients.  As for the usual $\theta$-functions, the multivariable
version obeys two quasiperiodicity properties in the complex plane.
These are, for any given $\mathbf{m} \in \mathbb{Z}^g$,
\begin{eqnarray}
\theta (\mathbf{z} + \pi \mathbf{m} | \mathbf{\Omega}) = \theta
(\mathbf{z} | \mathbf{\Omega}), \nonumber \\
\theta (\mathbf{z} + \pi \mathbf{\Omega} \mathbf{m} | \mathbf{\Omega})
= e^{-i\pi \mathbf{m}^t \mathbf{\Omega} \mathbf{m} - 2 i \mathbf{m}^t
\mathbf{z}} \theta (\mathbf{z} | \mathbf{\Omega}).
\end{eqnarray}

The limits of low or high temperature are obtained by taking 
$\det \mathbf {\Omega} \rightarrow \infty$ or  
$\det \mathbf {\Omega} \rightarrow 0$ respectively.
In the first case 
\begin{eqnarray}
\theta (\mathbf{z} | \mathbf{\Omega}) \rightarrow 1+ \sum_{i=1}^g e^{i \pi  {\mathbf {\Omega}}_{ii}} 
  2 \cos (2 \mathbf{z}_i )
\end{eqnarray}
For the high temperature limit we need to use  a generalization of the 
Poisson summation formula to 
$\mathbb{Z}^g$ given by 
\begin{eqnarray}
\theta (\mathbf{z} | \mathbf{\Omega}) = {\frac {1}{\sqrt {\det (-i {\mathbf{\Omega}})}}} 
\sum_{\mathbf{n} \in
\mathbb{Z}^g} e^{-i {\frac {1}{\pi}} (\mathbf{n}^t \pi + \mathbf{z}^t    ) \mathbf{\Omega^{-1}} (\mathbf{n} \pi+ \mathbf{z}  ) }.
\end{eqnarray}
The limit is then easily taken.

\section{Notes on the real-time perturbative formalism in finite-size
systems}
\label{apC} 

In order to compute thermodynamic physical quantities for a given
system at finite temperature, one usually uses the Matsubara formalism using 
imaginary time.  If one desires to obtain finite-temperature
time-dependent correlation functions, one then has to perform an
analytical continuation back to real times, an operation which, done
correctly, is often inconvenient.  

We wish here to discuss some details of the derivation of the
Matsubara formalism from first principles, and show that in our
geometry of interest, i.e. a one-dimensional finite-size system, there
are subtleties one has to bear in mind.  

We adopt the following logic, using a real-time formalism to start
with.  Let us suppose that at $t = -\infty$, the Hamiltonian of the
system is described by a known operatorial expression $H_0$.  We then
adiabatically turn on an additional piece $H_1$ with a modulation
function $f(t)$.  That is, $H(t) = H_0 + f(t) H_1$.  The fundamental
object that we need to consider is the density matrix $\rho (t)$,
which obeys the time-evolution equation
\begin{eqnarray}
i \frac{d}{dt} \rho(t) = \left[ H(t), \rho(t) \right].
\end{eqnarray}
Defining $\rho(t) = \rho_0 + \rho_1 (t)$ where $\rho_0 = e^{-\beta
H_0}/Z_0$, we can obtain a self-consistency equation for $\rho_1 (t)$:
\begin{eqnarray}
\rho_1 (t) = -i \int_{-\infty}^t dt_1 f(t_1) e^{-i H_0 (t-t_1)} \left
[ H_1, \rho_0 + \rho_1 (t_1) \right] e^{i H_0 (t-t_1)}
\end{eqnarray}
which we can solve as a series $\rho_1 (t) = \sum_{a = 1}^{\infty}
\rho_1^{(a)} (t) / Z_0$, each successive term containing one additional
time integral and commutator with the unperturbed density operator.

Let us consider the first correction.  For simplicity, we consider a
modulation function with periodicity $f(t - i\beta) = f(t)$, although
this is not a fundamental requirement.  Using the notation $H_1 (t) =
e^{i H_0 t} H_1 e^{-i H_0 t}$, we obtain after changing integration
variables and using the periodicity of $f(t)$
\begin{eqnarray}
\rho_1^{(1)} (t) = -i e^{-\beta H_0} \left[ \int_{-\infty -
i\beta}^{-i \beta} dt_1 - \int_{-\infty}^0 dt_1 \right] f(t_1 + t) H_1
(t_1). 
\end{eqnarray}
We can then combine the two integrals into one closed contour (using
$f(-\infty) = 0$ by construction) and a leftover Matsubara piece:
\begin{eqnarray}
\rho_1^{(1)} (t) = 2\pi e^{-\beta H_0} \sum_{s_j \in C (0)} Res(f(t_1
+ t) H_1 (t_1) - i e^{-\beta H_0} \int_0^{-i\beta} dt_1 f(t_1 + t) H_1
(t_1).
\end{eqnarray}
Let us suppose that we are in an infinite (and therefore aperiodic)
system.  If we choose to turn on our interaction $H_1$ at a very
remote point in the past as compared to any object we want to
calculate (say $t = 0$), then $f(t)$ will have poles only extremely
far away, and 
the first correction will yield expectation values with extremely
large time difference, which one can safely take to vanish.  Moreover, 
near $t = 0$, we suppose that $f(t)$ has reached its asymptotic
behaviour $f(t) \approx 1$.  The second term then becomes 
\begin{eqnarray}
\rho_{1,Mats}^{(1)} (t) = - e^{-\beta H_0} \int_0^{\beta} d \tau H_1 (-i\tau)
\end{eqnarray}
which is just the Matsubara contribution.

This logic, however, fails for a finite-size system described by an
initial $H_0$ containing no dissipation.  To see this, one simply has
to realize that in such a finite-size system, correlation functions are by
construction periodic in time, with a return time of the length of the
system divided by the appropriate velocity.  For example, one can take
as $f(t)$ a step function triggering the perturbation $H_1$ at time 
$t_0$.  In that case, the computation above gives
\begin{eqnarray}
\rho_1^{(1)} (t) = e^{-\beta H_0} \int_0^{\beta} d \tau \left[ H_1
(t_0 - i \tau) - H_1 (-i \tau) \right]
\end{eqnarray}
and no matter how far we put $t_0$ in the past, the first term never
vanishes, since all correlations in the system are periodic.
In all that we have done, we neglect the contribution from the first
term, which we presume ``die off'' for $t_0$ enough in the past.
For the second-order contribution, similar arguments yield
\begin{eqnarray}
\rho_1^{(2)} (t) = \frac{e^{-\beta H_0}}{2} \int_0^{\beta} d\tau_1
\int_0^{\beta} d\tau_2 f(t-i\tau_1) f(t-i\tau_2) T_{\tau}
\left[H_1(-i\tau_1) H_1(-i\tau_2) \right],
\end{eqnarray}
(where $T_{\tau}$ is the imaginary-time ordering operator)
which we use in the text with the same simplification as used in the
first-order term, i.e. replace $f(t-i\tau) =1$ in the form above.

\acknowledgments
D.S. acknowledges support from the Della Riccia Foundation (Florence, Italy) 
and from the Isaac Newton Institute for Mathematical Sciences (Cambridge, UK)
(Junior Member affiliation).
 A. L. acknowledges support from CONICET (Argentina), and EPSRC (U.K.).

\end{document}